\documentclass[
aps,
prd,
tightenlines,
superscriptaddress,
nofootinbib,
floatfix,
groupedaddress,
nofootinbib,
twocolumn,
amsmath,
amsfonts,
amssymb]
{revtex4}
\usepackage{notoccite}
\usepackage{graphicx,
longtable,
color
}

\usepackage{xspace}
\usepackage[normalem]{ulem}
\newcommand{\nua}[1]{\ensuremath{\rlap{\kern-1.0pt\ensuremath{\overset{\scriptscriptstyle(-)}{\phantom{\nu}}}}{\ensuremath{{\nu}_{#1}}}}\xspace}

\begin{document}
\title{Joint short- and long-baseline constraints on light sterile neutrinos}%
\author{		F.~Capozzi}
\affiliation{ 	Dipartimento di Fisica e Astronomia ``Galileo Galilei'', Universit\`a di Padova, Via F.\ Marzolo 8, I-35131 Padova, Italy}
\affiliation{		Istituto Nazionale di Fisica Nucleare (INFN), Sezione di Padova, Via F.\ Marzolo 8, I-35131 Padova, Italy}

\author{		C. Giunti}
\affiliation{   	Istituto Nazionale di Fisica Nucleare (INFN), Sezione di Torino, Via P.\ Giuria 1, I-10125 Torino, Italy}

\author{		M. Laveder}
\affiliation{ 	Dipartimento di Fisica e Astronomia ``Galileo Galilei'', Universit\`a di Padova, Via F.\ Marzolo 8, I-35131 Padova, Italy}
\affiliation{		Istituto Nazionale di Fisica Nucleare (INFN), Sezione di Padova, Via F.\ Marzolo 8, I-35131 Padova, Italy}

\author{		A.~Palazzo}
\affiliation{ 	Dipartimento Interateneo di Fisica ``Michelangelo Merlin,'' Via Amendola 173, 70126 Bari, Italy}
\affiliation{ 	Istituto Nazionale di Fisica Nucleare, Sezione di Bari, Via Orabona 4, 70126 Bari, Italy}


\begin{abstract}

Recent studies have evidenced that long-baseline (LBL) experiments are sensitive to the extra CP-phases involved with light sterile neutrinos, whose existence is suggested by several anomalous short-baseline (SBL) results. We show that, within the 3+1 scheme,  the combination of the existing SBL data with the LBL results coming from the two currently running experiments NO$\nu$A and T2K, enables us to simultaneously constrain two active-sterile mixing angles $\theta_{14}$ and $\theta_{24}$ and two CP-phases $\delta_{13} \equiv \delta$ and $\delta_{14}$, albeit the information on the second CP-phase is still weak at the moment.  The two mixing angles are basically determined by the SBL data, while the two CP-phases are constrained by the LBL experiments, once the information coming from the SBL setups is taken into account. We also assess the robustness/fragility of the estimates of the standard 3-flavor parameters in the more general 3+1 scheme. To this regard we find that: i) the indication of CP-violation found in the 3-flavor analyses persists also in the 3+1 scheme, with $\delta_{13} \equiv \delta$ having still its best fit value around $-\pi/2$; ii) the 3-flavor weak hint in favor of the normal hierarchy becomes even less significant when sterile neutrinos come into play; iii) the weak indication of non-maximal $\theta_{23}$  (driven by NO$\nu$A disappearance data) persists in the 3+1 scheme, where
maximal mixing is disfavored at almost the 90\% C.L. in both normal and inverted mass hierarchy; iv) the preference in favor of one of the two octants of $\theta_{23}$ found in the 3-flavor framework (higher octant for inverted mass hierarchy) is completely washed out in the 3+1 scheme.

\end{abstract}
\maketitle

\section{Introduction}

The massive nature of neutrinos and their mixing have been established by a plethora
of experiments performed in the last two decades with natural and artificial neutrino sources.
The 3-flavor paradigm has been gradually recognized as the sole framework able 
to account for all the observations performed at baselines longer than $\sim 100$ meters.
In contrast, the same scheme is not able to explain a series of results recorded
at shorter distances dubbed as short-baseline (SBL) anomalies. One possible 
explanation of the SBL anomalies is provided by a flavor oscillation process mediated 
by new hypothetical light sterile neutrino states, albeit this hypothesis encounters
some difficulties in explaining simultaneously all the existing data sets. In particular,
a non-negligible statistical tension emerges when one compares the (positive) 
$\nu_\mu \to \nu_e$ appearance signals with the joint disappearance results
from the (positive) searches in the $\nu_e \to \nu_e$ channel and the (negative)
ones in the $\nu_\mu \to \nu_\mu$ channel.
   
In the so-called 3+1 scheme, only one new (essentially sterile) mass eigenstate is introduced,
with a squared-mass splitting of the order of $\Delta m^2_{\text{SBL}} \sim 1\,{\mathrm {eV}}^2$
with respect to the three standard neutrinos. In such a scheme 
the active-sterile admixture is supposed to be small but large enough to explain the anomalies.
The 3+1 scheme predicts by construction sizable effects at the short distances,
where the oscillating factor $\Delta_{\text{SBL}} = \Delta m^2_{\text{SBL}} L/4E$ ($L$ being the baseline and $E$ the neutrino energy) is of order one.  A rich program of new SBL experiments is underway
with the purpose of detecting the smoking gun of active-sterile neutrino oscillations,
i.e., the characteristic $L/E$ dependence of the events rate.

After a hypothetical discovery of a sterile neutrino at SBL experiments, the full exploration
of the properties of these particles would need other types of experimental setups.
In particular, the SBL experiments would not be able to provide any information on the CP-phases involved in the 3+1 scheme.
In fact, the manifestation of CP-violation (CPV) is intimately related to the interference of two 
distinct frequencies. At the SBL setups only one frequency is observable (the new one),
while the other two (atmospheric and solar) are undetectable. Therefore, the 
SBL searches are blind to the CP-phases involved in the 3+1 scheme.%
 \footnote{In the $3+N_{s}$ schemes with $N_{s} >1$,  CPV could be observed at SBL experiments. 
 However, these setups can probe only a limited number of all the CP phases involved in the model.
 In contrast,  LBL experiments have access to all such phases. For example, in the 3+2 scheme, 
 the SBL experiments are sensitive only to one CP-phase over a total of five CP-phases. }
 
In the standard 3-flavor framework the CPV is encoded by the CP-phase $\delta$
which enters the leptonic mixing matrix. The 3-flavor CPV searches are performed at the long-baseline (LBL)
experiments, which can observe the $\nu_\mu \to \nu_e$ transition probability in both
the neutrino and antineutrino channels. These setups are designed to maximize the 
amplitude of the interference between the solar and atmospheric
oscillations, which embodies a dependency on the CP-phase $\delta$. 
As a matter of fact, we have already some intriguing indications on $\delta$ coming
from the partial~\cite{palazzo_klop,palazzo_t2k_nova} (including only LBL data)
and global~\cite{global_analysis_2016,marrone_neutrino_2016,Esteban:2016qun,Forero:2014bxa} analyses, which all 
point towards nearly maximal CPV with $\delta \sim -\pi/2$. This trend has been
recently corroborated by the latest data released by NO$\nu$A~\cite{nova_neutrino_2016} and T2K~\cite{t2k_ichep_2016} at the Neutrino and ICHEP 2016 conferences.

As first evidenced in Ref.~\cite{palazzo_klop}, in the presence of light sterile neutrinos, 
the $\nu_\mu \to \nu_e$ transition probability probed at the LBL facilities
acquires a new interference term, arising from the interference between the atmospheric frequency
and the new large frequency related to the sterile state. Although the fast
oscillations driven by the new frequency are completely averaged out by the finite 
energy resolution of the detector, nonetheless, they can leave their footprints in the transition probability.
This renders the LBL experiments sensitive also to the extra CP-phases involved in the 3+1 scheme. 
The recent 4-flavor analyses~\cite{palazzo_klop,palazzo_t2k_nova} of the data from NO$\nu$A and T2K have clearly shown
that these two experiments are already sensitive to one of the new CP-phases
provided that the active-sterile mixing angles are fixed at their best fit values determined by 
the SBL 3+1 fits~\cite{Giunti:2013aea,kopp_sterile,Collin:2016rao,Collin:2016aqd}. In addition, the prospective study performed
in~\cite{palazzo_t2k_nova_sterile} has shown that the 
sensitivity to the extra CP-phases is expected to improve when NO$\nu$A and 
T2K will reach their full exposures, and will further increase in the next-generation 
experiment DUNE~\cite{palazzo_dune_sterile}.

In this work, we stick to the real data and take a step forward with respect to the existing works.
Instead of fixing the active-sterile mixing angles at their best fit values, we here incorporate
a full analysis of all the existing SBL data in combination with the LBL results. In this way we 
are able to simultaneously constrain the two active-sterile mixing angles $\theta_{14}$ and $\theta_{24}$
and the two CP-phases $\delta_{13} \equiv \delta$ and $\delta_{14}$.
The estimates of the two new mixing angles will be basically determined by the SBL data, while
those of the two CP-phases will derive from the LBL experiments, once the information from the 
SBL setups is considered. In this work, we also assess the robustness/fragility of the estimates of the 
standard 3-flavor parameters in the more general 3+1 scheme, paying particular attention
to the most important properties currently under scrutiny: the mass hierarchy, the CP phase $\delta$,
and the atmospheric mixing angle $\theta_{23}$. 

The rest of the paper is organized as follows.
In Sec. II we briefly introduce the theoretical 3+1 framework.  In Sec. III we recall the basic features
of the flavor oscillations at short baselines. Section IV deals with the 4-flavor transition probability  
relevant for the LBL setups. In Sec. V we list the data used in the simulations and describe
the details of their numerical analysis. In Sec. VI we present and discuss the results of the analysis. 
Finally, in Sec. VII we draw our conclusions.

\section{Theoretical framework}

In the 3+1 scheme, the flavor ($\nu_e,\nu_\mu, \nu_\tau, \nu_s$) 
and mass eigenstates ($\nu_1, \nu_2, \nu_3, \nu_4$) are related 
through a $4\times4$ mixing matrix, which we parametrize as 
\begin{equation}
\label{eq:U}
U =   \tilde R_{34}  R_{24} \tilde R_{14} R_{23} \tilde R_{13} R_{12}\,, 
\end{equation} 
where $R_{ij}$ ($\tilde R_{ij}$) is a real (complex) $4\times4$ rotation in the ($i,j$) plane, 
which contains the $2\times2$ matrix 
\begin{eqnarray}
\label{eq:R_ij_2dim}
     R^{2\times2}_{ij} =
    \begin{pmatrix}
         c_{ij} &  s_{ij}  \\
         - s_{ij}  &  c_{ij}
    \end{pmatrix}
\,\,\,\,\,\,\,   
     \tilde R^{2\times2}_{ij} =
    \begin{pmatrix}
         c_{ij} &  \tilde s_{ij}  \\
         - \tilde s_{ij}^*  &  c_{ij}
    \end{pmatrix}
\,,    
\end{eqnarray}
in the $(i,j)$ sub-block, where we have introduced the definitions
\begin{eqnarray}
 c_{ij} \equiv \cos \theta_{ij} \qquad s_{ij} \equiv \sin \theta_{ij}\qquad  \tilde s_{ij} \equiv s_{ij} e^{-i\delta_{ij}}.
\end{eqnarray}
The parameterization in Eq.~(\ref{eq:U}) has the following features: i) When all the three mixing
angles involving the fourth state vanish $(\theta_{14} = \theta_{24} = \theta_{34} =0)$ 
one recovers the 3-flavor matrix in its standard parameterization.
ii) For small values of the mixing angles involving the fourth mass eigenstate, 
it is  $|U_{e4}|^2 = s^2_{14}$,  $|U_{\mu4}|^2  \simeq s^2_{24}$ and 
$|U_{\tau4}|^2 \simeq s^2_{34}$, with an immediate physical 
interpretation of the new mixing angles.  iii) With the leftmost positioning of the matrix $\tilde R_{34}$,
in vacuum, the LBL $\nu_{\mu} \to \nu_{e}$  transition probability is 
independent of $\theta_{34}$ and of the associated CP-phase $\delta_{34}$.

\section{Flavor conversion at short baselines}

Short-baseline experiments are sensitive only to the oscillations
generated by the new squared-mass difference
$\Delta m^2_{\text{SBL}} \sim 1\,{\mathrm {eV}}^2$,
which in the 3+1 framework is
$\Delta m^2_{\text{SBL}} = \Delta m^2_{41} \simeq \Delta m^2_{42} \simeq \Delta m^2_{43} $,
taking into account that
the solar squared-mass difference
$\Delta m^2_{\text{SOL}} = \Delta m^2_{21} \approx 7.4 \times 10^{-5} \, \text{eV}^2$
and the atmospheric squared-mass difference
$\Delta m^2_{\text{ATM}} = |\Delta m^2_{31}| \simeq |\Delta m^2_{32}| \approx 2.5 \times 10^{-3} \, \text{eV}^2$
are much smaller
(we use the notation $\Delta m^2_{jk} = m_{j}^2 - m_{k}^2$).

The effective oscillation probabilities of the flavor neutrinos in short-baseline experiments
are given by~\cite{Bilenky:1996rw}
\begin{equation}
P_{\alpha\beta}^{(\text{SBL})}
\simeq
\left|
\delta_{\alpha\beta}
-
\sin^2 2\theta_{\alpha\beta}
\sin^{2}\!\left( \frac{\Delta{m}^2_{41}L}{4E} \right)
\right|,
\label{eq:sblosc}
\end{equation}
where $\alpha,\beta =e,\mu,\tau,s$,
$L$ is the source-detector distance and $E$ is the neutrino energy.
The short-baseline oscillation amplitudes depend only on the absolute values of the
elements in the fourth column of the mixing matrix:
\begin{equation}
\sin^2 2\theta_{\alpha\beta}
=
4
|U_{\alpha 4}|^2
\left| \delta_{\alpha\beta} -  |U_{\beta 4}|^2 \right|
.
\label{amp3p1}
\end{equation}
Hence, the transition probabilities of neutrinos and antineutrinos are equal
and it is not possible to measure any CPV effect generated by the complex phases 
in the mixing matrix in short-baseline experiments.

The short-baseline anomalies in favor of the existence of
active-sterile neutrino oscillations are:

\begin{enumerate}

\renewcommand{\labelenumi}{\theenumi.}
\renewcommand{\theenumi}{\arabic{enumi}}

\item
The
LSND observation of an excess of
$\bar\nu_{e}$-induced events
in a $\bar\nu_{\mu}$ beam~\cite{Athanassopoulos:1995iw,Aguilar:2001ty}.

\item
The Gallium neutrino anomaly~\cite{Abdurashitov:2005tb,Laveder:2007zz,Giunti:2006bj,Giunti:2010zu,Giunti:2012tn},
consisting in the disappearance of $\nu_{e}$
measured in the
Gallium radioactive source experiments
GALLEX~\cite{Kaether:2010ag}
and
SAGE~\cite{Abdurashitov:2009tn}.

\item
The reactor antineutrino anomaly~\cite{Mention:2011rk},
which is a deficit of the rate of $\bar\nu_{e}$ observed in several
reactor neutrino experiments
in comparison with that expected from the calculation of
the reactor neutrino fluxes~\cite{Mueller:2011nm,Huber:2011wv}.

\end{enumerate}

\section{Flavor conversion at long baselines}

Let us now come to the transition probability relevant for the LBL experiments T2K and NO$\nu$A.
In Ref.~\cite{palazzo_klop}, it has been shown that the probability can be written
as the sum of three terms
\begin{eqnarray}
\label{eq:Pme_4nu_3_terms}
P^{4\nu}_{\mu e}  \simeq  P^{\rm{ATM}} + P^{\rm {INT}}_{\rm I}+   P^{\rm {INT}}_{\rm II}\,.
\end{eqnarray}
The first term represents the positive definite atmospheric transition probability, which can be
expressed as 
\begin{eqnarray}
\label{eq:Pme_atm}
 &\!\! \!\! \!\! \!\! \!\! \!\! \!\!  P^{\rm {ATM}} &\!\! \simeq\,  4 s_{23}^2 s^2_{13}  \sin^2{\Delta}\,,
\end{eqnarray}
where $\Delta \equiv  \Delta m^2_{31}L/4E$ is the atmospheric oscillating frequency.
The second term is related to the interference between the oscillations driven by the
solar and atmospheric squared-mass splittings. This term, apart from higher order corrections,
coincides with the standard interference term, which makes the transition probability sensitive
to the CP-phase $\delta \equiv \delta_{13}$. It can be written as
\begin{eqnarray}
 \label{eq:Pme_int_1}
 &\!\! \!\! \!\! \!\! \!\! \!\! \!\! \!\! P^{\rm {INT}}_{\rm I} &\!\!  \simeq\,   8 s_{13} s_{12} c_{12} s_{23} c_{23} (\alpha \Delta)\sin \Delta \cos({\Delta + \delta_{13}})\,.
\end{eqnarray}
The third term is due to 4-flavor effects and is driven by the interference between the atmospheric
frequency and the new large frequency introduced by the fourth mass eigenstate. It
takes the form
\begin{eqnarray}
 \label{eq:Pme_int_2}
 &\!\! \!\! \!\! \!\! \!\! \!\! \!\! \!\! P^{\rm {INT}}_{\rm II} &\!\!  \simeq\,   4 s_{14} s_{24} s_{13} s_{23} \sin\Delta \sin (\Delta + \delta_{13} - \delta_{14})\,.
\end{eqnarray}
This term does not depend on $\Delta m^2_{41}$ because the fast oscillations are
averaged by the finite resolution of the detector. The transition probability
depends on the three small mixing angles $s_{13}, s_{14}, s_{24} \simeq 0.15$,
which can be all assumed to be of the same order $\epsilon$. Another small quantity is
the ratio of the solar and atmospheric squared-mass splittings  $\alpha \equiv \Delta m^2_{12}/ \Delta m^2_{13} \simeq \pm 0.03$, 
which is of order $\epsilon^2$. Remarkably, for values of the mixing angles indicated
by the current global 3-flavor analyses~\cite{global_analysis_2016,marrone_neutrino_2016,Esteban:2016qun,Forero:2014bxa}
(for $\theta_{13}$) and the $3+1$ fits~\cite{Giunti:2013aea,kopp_sterile,Collin:2016rao,Collin:2016aqd} (for $\theta_{14}$ and $\theta_{24}$), the size of
the new (atmospheric-sterile) interference term is basically identical to that of the standard 
(solar-atmospheric) interference term~\cite{palazzo_t2k_nova}.  This implies that T2K and NO$\nu$A 
are sensitive to both  CP-phases $\delta_{13}$ and $\delta_{14}$.

Finally, we mention that the matter effects slightly modify the transition probability, 
leaving unaltered its decomposition in the  sum of three contributions. We refer the reader
to~\cite{palazzo_klop} for a detailed treatment of matter effects in the 3+1 scheme. Here, we just recall
that they introduce a dependency on the dimensionless quantity
\begin{equation}
\label{eq:v}\,
v = \frac{2VE}{\Delta m^2_{31}}\,,
\end{equation}
where 
\begin{equation}
\label{eq:potential}
V = \sqrt 2 G_F N_e\,
\end{equation}
is the constant matter potential along the neutrino trajectory in the Earth crust.  
We have $v\simeq 0.05$ in T2K and $v\simeq 0.17$ in NO$\nu$A at the energy corresponding
to the first oscillation maximum ($E \simeq 0.6$ GeV in T2K, $E \simeq 2$ GeV in NO$\nu$A). 
Therefore, the matter effects are appreciable only in NO$\nu$A and confer to this
experiment an enhanced sensitivity to the neutrino mass hierarchy.

\section{Data used and details of the analysis}

For the determination of the two active-sterile mixing angles $\theta_{14}$ and $\theta_{24}$
we use the update of the analysis in Ref.~\cite{Gariazzo:2015rra}
presented in Ref.~\cite{Giunti:2016oan}.
We considered the data of the following three groups of experiments:

\begin{enumerate}

\renewcommand{\labelenumi}{(\theenumi)}
\renewcommand{\theenumi}{\Alph{enumi}}

\item
The
$\nua{\mu}\to\nua{e}$
appearance data of the
LSND~\cite{Aguilar:2001ty},
MiniBooNE~\cite{AguilarArevalo:2008rc,Aguilar-Arevalo:2013pmq},
BNL-E776~\cite{Borodovsky:1992pn},
KARMEN~\cite{Armbruster:2002mp},
NOMAD~\cite{Astier:2003gs},
ICARUS~\cite{Antonello:2013gut}
and
OPERA~\cite{Agafonova:2013xsk}
experiments. The two last ones have been treated following the approach described in~\cite{Palazzo:2015wea}.
We did not consider the anomalous low-energy bins of the MiniBooNE experiment~\cite{AguilarArevalo:2008rc,Aguilar-Arevalo:2013pmq},
according to the ``pragmatic'' approach advocated in Ref.~\cite{Giunti:2013aea}.

\item
The following
$\nua{e}$
disappearance data:
1)
the data of the
Bugey-4~\cite{Declais:1994ma},
ROVNO91~\cite{Kuvshinnikov:1990ry},
Bugey-3~\cite{Declais:1995su},
Gosgen~\cite{Zacek:1986cu},
ILL~\cite{Hoummada:1995zz},
Krasnoyarsk~\cite{Vidyakin:1990iz},
Rovno88 ~\cite{Afonin:1988gx},
SRP~\cite{Greenwood:1996pb},
Chooz~\cite{Apollonio:2002gd},
Palo Verde~\cite{Boehm:2001ik},
Double Chooz~\cite{Abe:2014bwa}, and
Daya Bay~\cite{An:2015nua}
reactor antineutrino experiments
with the new theoretical fluxes~\cite{Mueller:2011nm,Huber:2011wv,Mention:2011rk,Abazajian:2012ys};
2)
the data of the
GALLEX~\cite{Kaether:2010ag}
and
SAGE~\cite{Abdurashitov:2009tn}
Gallium radioactive source experiments
with the statistical method discussed in Ref.~\cite{Giunti:2010zu},
considering the recent
${}^{71}\text{Ga}({}^{3}\text{He},{}^{3}\text{H}){}^{71}\text{Ge}$
cross section measurement in Ref.~\cite{Frekers:2011zz};
3)
the solar neutrino constraint on $\sin^{2}2\theta_{ee}$~\cite{Giunti:2009xz,Palazzo:2011rj,Palazzo:2012yf,Giunti:2012tn,Palazzo:2013me};
4)
the
KARMEN~\cite{Armbruster:1998uk}
and
LSND~\cite{Auerbach:2001hz}
$\nu_{e} + {}^{12}\text{C} \to {}^{12}\text{N}_{\text{g.s.}} + e^{-}$
scattering data~\cite{Conrad:2011ce},
with the method discussed in Ref.~\cite{Giunti:2011cp}.

\item
The constraints on
$\nua{\mu}$
disappearance obtained from
the data of the
CDHSW experiment~\cite{Dydak:1983zq},
from the analysis~\cite{Maltoni:2007zf} of
the data of
atmospheric neutrino oscillation experiments,
from the analysis~\cite{Hernandez:2011rs,Giunti:2011hn} of the
MINOS neutral-current data~\cite{Adamson:2011ku}
and from the analysis of the
SciBooNE-MiniBooNE data
neutrino~\cite{Mahn:2011ea} and antineutrino~\cite{Cheng:2012yy} data.
We have not included the IceCube results recently reported in~\cite{TheIceCube:2016oqi}.
However, this has no impact in our results, because, as already noted in~\cite{Giunti:2016oan},
these data modify the upper bounds on $\theta_{24}$ only for values of $\Delta m^2_{41}$ 
which are lower then $\sim$1\,eV$^2$. This conclusion is corroborated by the numerical analysis
performed in~\cite{Collin:2016aqd}.

\end{enumerate}

Concerning the LBL experiments, we use the preliminary data released at the 
Neutrino 2016 and ICHEP 2016 conferences by the NO$\nu$A~\cite{nova_neutrino_2016} 
and T2K~\cite{t2k_ichep_2016} collaborations, considering the neutrino and
antineutrino datasets, and including both the appearance and disappearance 
channels. In order to calculate the theoretical expectation for the total number of events
and their binned spectra in the reconstructed neutrino energy, we use the 
software GLoBES~\cite{globes_1,globes_2}. As input information
we use the unoscillated $\nu_{\mu}$ and $\overline{\nu}_{\mu}$  fluxes extrapolated 
at the far detector from Ref.~\cite{t2k_flux,t2k_antinu_flux} for T2K 
and from Ref.~\cite{nova_thesis} for NO$\nu$A.
The analysis for the appearance channel is performed 
using the total rate information as in~\cite{palazzo_klop,palazzo_t2k_nova}, which presents
very small differences with respect to the analysis done using the full energy spectrum. This is
due to three factors: i) the off-axis configuration of NO$\nu$A and T2K, which leads to a narrow energy spectrum
peaked around the first oscillation maximum; ii) the limited statistics currently available in the
appearance channel both in NO$\nu$A and  T2K; iii) the smearing 
induced by the finite energy resolution of the far detectors. Differently, for the disappearance channel we
perform a full spectral analysis of the far detector event distribution, since in this case the energy
information has a crucial role.

In the standard 3-flavor case, the free parameters in the analysis 
are the atmospheric mass splitting $\Delta m^2_{32}$, the two mixing angles $\theta_{13}$, $\theta_{23}$ 
and the CP-phase $\delta_{13}$. In the 4-flavor analysis, in addition, we consider 
as free parameters $\Delta m^2_{41}$, $\theta_{14}$, $\theta_{24}$ and the CP-phase $\delta_{14}$.
We fix $\theta_{34}=0$, beacause the perturbations induced by non-zero $\theta_{34}$ are
very small in T2K and NO$\nu$A. We have explicitly checked numerically that for non-zero values 
of $\theta_{34}$ currently allowed by data, the oscillation probabilities in both the appearance and disappearance
channels are almost indistinguishable from those calculated with $\theta_{34}=0$.
Finally, we mention that both in the 3-flavor and 4-flavor analyses we fix the solar mass-mixing 
parameters at their best fit values obtained in the global 3-flavor analysis~\cite{global_analysis_2016}.

As pointed out in Ref.~\cite{palazzo_klop}, in the 4-flavor scenario, the analysis has
to deal with the fact that the near detectors in the long-baseline experiments T2K and NO$\nu$A
are sensitive to the oscillations induced by the extra sterile neutrino. The neutrino fluxes used in
the standard analysis are constrained with the information extracted from the near detectors under
the assumption of no oscillation at short baselines, which is true only for three flavors. 
With the addition of an extra neutrino with $\Delta m^2_{41}\sim O$(1 eV$^{2}$), the survival probability for
$\nu_{\mu}$ at the near detector can be approximated as
\begin{equation}
P^{4\nu,\text{ND}}_{\mu\mu}\simeq 1-4\sin^2\theta_{24}\sin^2\left(\frac{\Delta m^2_{41}L}{4E}\right)\ .
\label{P_mumu_near}
\end{equation}
Therefore, a suppression of the fluxes that
depends on the parameters $\Delta m^2_{41}$, $\theta_{24}$ and on the
the energy is expected. A precision analysis of the LBL data 
in the 3+1 scheme would require the simultaneous treatment of the
near and far detector. However, the spectrum of events expected 
at the near detector is problematic to reproduce, since many details
are not accessible from outside the collaborations. To circumvent
this problem we have incorporated the effects of the oscillation at the
near detector using the following approximate procedure.
We have corrected the expected distribution of events at the far detector multiplying it 
by the energy dependent factor $1/P_{\mu\mu}^{4\nu,\text{ND}}$, taking its averaged value 
in each energy bin. In this way, we approximately untie the far detector fluxes from
their dependency on the oscillations occurred at the near detector. We have checked that these
corrections introduce small modifications on our final results.
Hence, our approximate approach is justified. However,
we stress that a more detailed analysis in the 3+1 scheme should incorporate
the simultaneous fit of the native neutrino fluxes with the near and far detector data
for varying values of the parameters $\Delta m^2_{41}$ and $\theta_{24}$.
At the moment, this is possible only from inside the experimental collaborations.

As a separate analysis, we constrain the value of $\theta_{13}$ using the far-to-near ratios of
the reactor $\theta_{13}$-sensitive experiments Daya-Bay and RENO. Their data are analyzed 
using the total rate information following the approach described in Ref.~\cite{palazzo_reactors}.
For both experiments we have used the latest data~\cite{daya_bay_neutrino_2016,reno_neutrino_2016}
based, respectively, on 1230 live days (Daya Bay) and 500 live days (RENO),
recently released at the Neutrino 2016 conference. Since the fast oscillations induced by 
$\Delta m^2_{14}$ are averaged out at both near and far detector sites,
the far-to-near ratios of Daya Bay and RENO are not sensitive to 4-flavor effects.
Therefore, the estimate of $\theta_{13}$ is identical in the 3-flavor and 3+1 schemes.
 
\begin{figure}[t!]
\vspace*{0.0cm}
\hspace*{-0.1cm}
\includegraphics[width=9.5 cm]{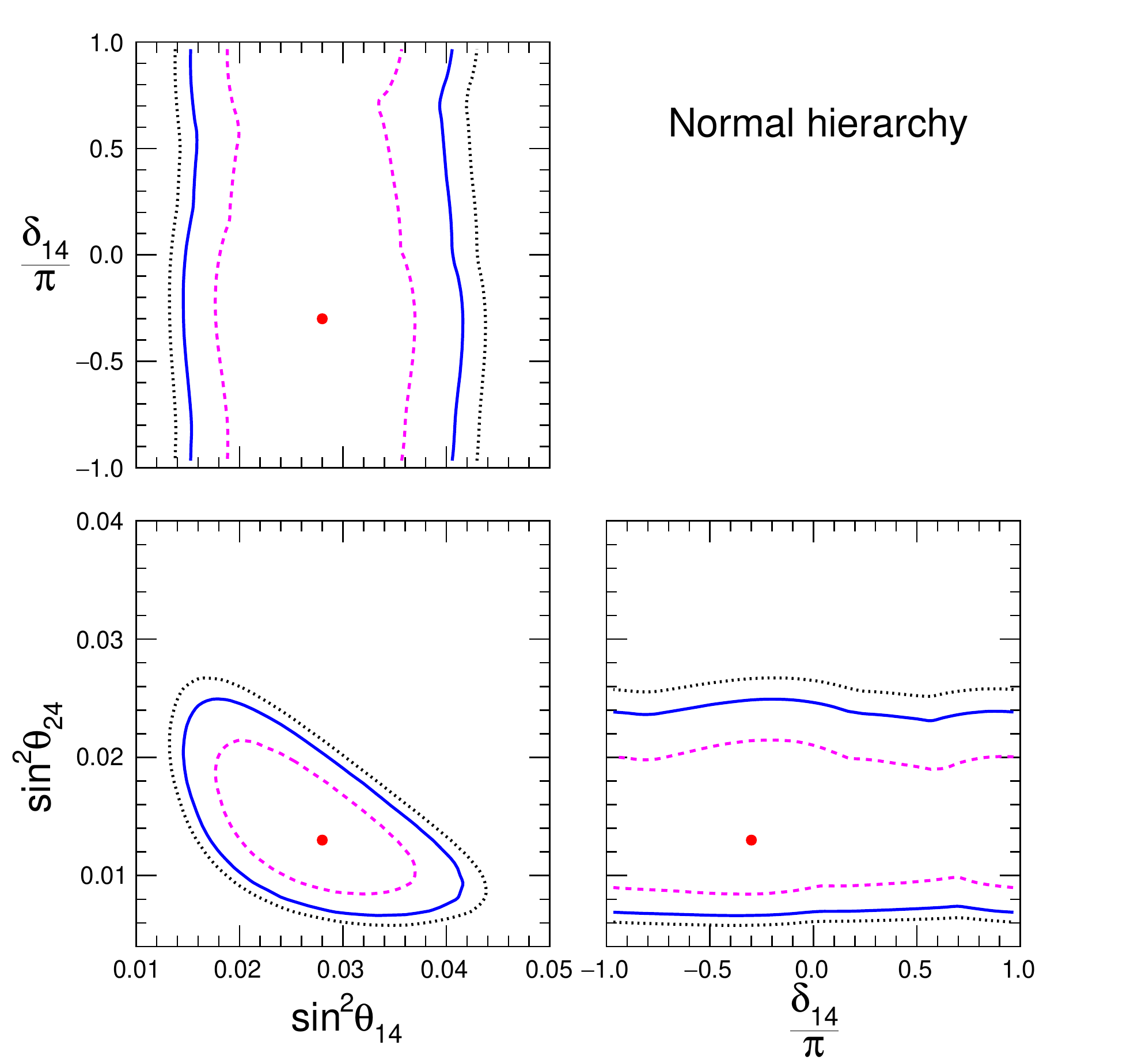}
\vspace*{-0.6cm}
\caption{Regions allowed by the combination of the SBL and LBL data (T2K and NO$\nu$A)
together with the $\theta_{13}$-sensitive reactor results for the NH case.
The left-bottom panel reports the projection on the plane of the two mixing
angles $(\theta_{14}, \theta_{24})$. The other two panels display the 
constraints in the plane formed by each one of these two mixing angles and the 
new CP-phase $\delta_{14}$. The confidence levels correspond to 68\%, 
90\% and 95\% for 2 d.o.f ($\Delta\chi^2=2.3,\ 4.6,\ 6.0$), and the best-fit points 
are marked with a red point.
\label{fig:3pan_sbl_lbl_NH}}
\end{figure}  
 
\section{Numerical Results}

\subsection{Constraints on the new mixing angles ($\theta_{14}$, $\theta_{24}$) and the new 
CP-phase $\delta_{14}$}

\begin{figure}[t!]
\vspace*{0.0cm}
\hspace*{-0.1cm}
\includegraphics[width=9.5 cm]{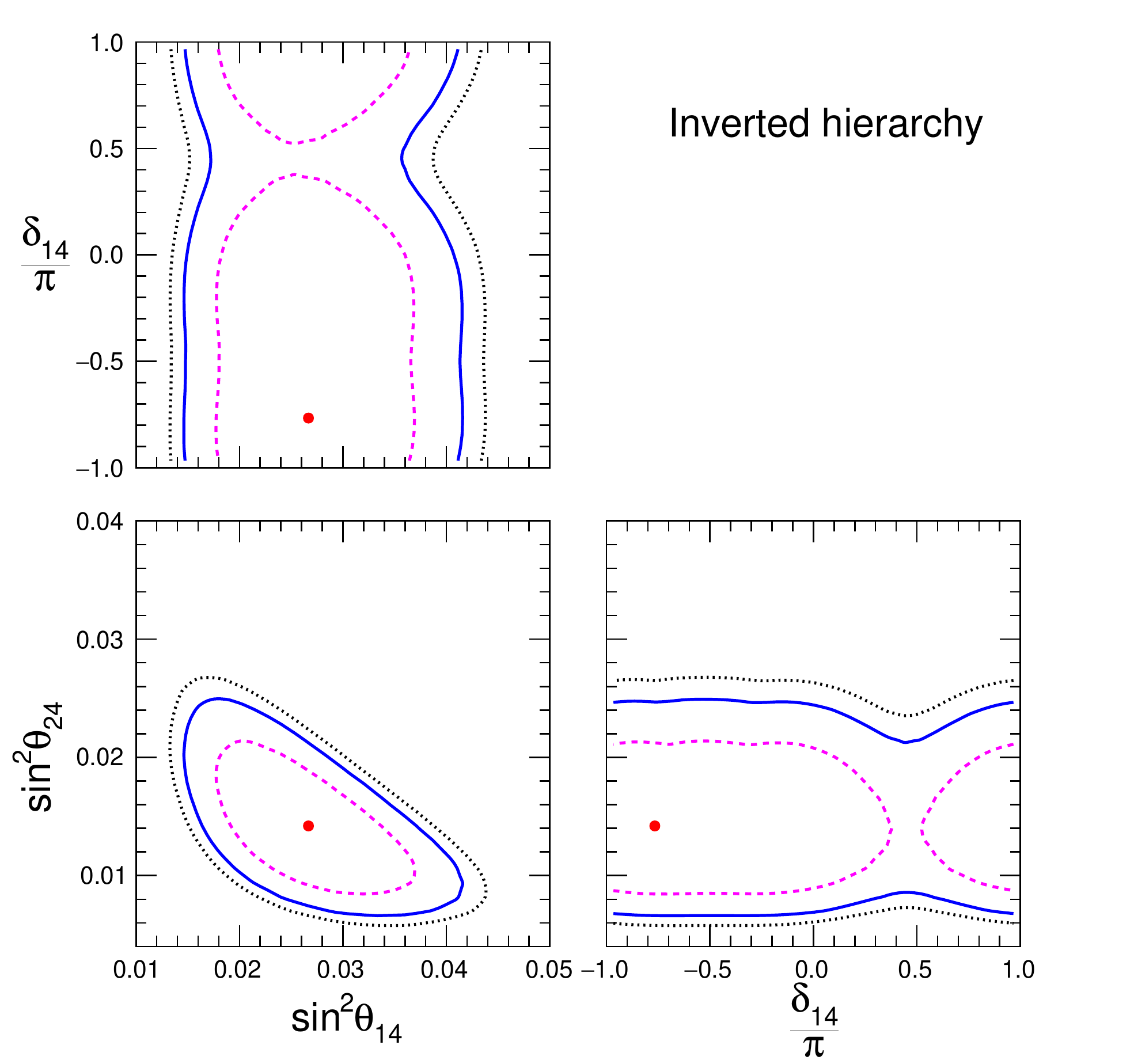}
\vspace*{-0.6cm}
\caption{Regions allowed by the combination of the SBL and LBL data (T2K and NO$\nu$A)
together with the $\theta_{13}$-sensitive reactor results for the IH case.
The left-bottom panel reports the projection on the plane of the two mixing
angles $(\theta_{14}, \theta_{24})$. The other two panels display the 
constraints in the plane formed by each one of these two mixing angles and the 
new CP-phase $\delta_{14}$. The confidence levels correspond to 68\%, 
90\% and 95\% for 2 d.o.f ($\Delta\chi^2=2.3,\ 4.6,\ 6.0$), and the best-fit points 
are marked with a red point.
\label{fig:3pan_sbl_lbl_IH}}
\end{figure}

Figure~\ref{fig:3pan_sbl_lbl_NH} and~\ref{fig:3pan_sbl_lbl_IH} represent the 
bidimensional projections of the $\Delta \chi^2$ for normal hierarchy (NH) and 
inverted hierarchy (IH) in the planes  [$\sin^2\theta_{14},\delta_{14}$], [$\sin^2\theta_{14},\sin^2\theta_{24}$] and 
[$\delta_{14},\sin^2\theta_{24}$] for the top left, bottom left and bottom right 
panels respectively. The three contours are drawn for $\Delta\chi^2=2.3,\,\, 4.6,\,\, 6.0$,
corresponding to 68\%, 90\% and 95\%  for 2 d.o.f.
 The allowed regions in the [$\sin^2\theta_{14},\sin^2\theta_{24}$] plane are almost the same
of those (not shown) that we obtain from the fit of the SBL data taken alone. This finding can be 
understood by observing that the SBL experiments currently dominate over the LBL ones
in the determination of the two new mixing angles. We find that the overall goodness of fit is 
satisfactory (${\mathrm {GoF}} = 24\%$), while the parameter goodness of fit (see~\cite{Maltoni:2003cu} for its definition), 
which measures the statistical compatibility between the (discordant) appearance and
disappearance data sets, is lower (${\mathrm {GoF}} = 7\%$). This implies that even if the 
closed contours presented for the two new mixing angles $\theta_{14}$ and $\theta_{24}$
exclude zero with high significance (more than six standard deviations), one cannot
naively interpret this circumstance as an evidence for sterile neutrinos. In addition,
we mention that light sterile neutrinos, unless dressed with new properties, 
are in strong tension with cosmological data (see for example~\cite{Hannestad:2012ky,Dasgupta:2013zpn,Hannestad:2013ana,Saviano:2013ktj,Archidiacono:2016kkh}). 

The preferred values  of $\sin^2\theta_{14}$ and $\sin^2\theta_{24}$ lie in the range [0.01-0.03], 
which means that the new mixing angles $\theta_{14}$ and $\theta_{24}$ are of the same order 
of magnitude of the standard mixing angle $\theta_{13}$ (we recall that $\sin^2\theta_{13} \simeq 0.025$). 
A quick estimate of the amplitude of the new interference term in Eq.~(\ref{eq:Pme_int_2}) reveals
that its size is similar to that of the standard interference term in Eq.~(\ref{eq:Pme_int_1}).
Therefore, it is quite natural to expect that the LBL data will posses some sensitivity to the new CP-phase  $\delta_{14}$.
This qualitative conclusion is validated by our numerical results displayed in the top left and bottom right
panels of  Figs.~\ref{fig:3pan_sbl_lbl_NH} and \ref{fig:3pan_sbl_lbl_IH}.
It is important to observe that  the input from the SBL experiments is essential for
the extraction of the information on $\delta_{14}$ from the LBL setups, 
since these last ones have a very scarce sensitivity to $\theta_{14}$ and $\theta_{24}$, and therefore
are unable to constrain the amplitude of the new interference term in Eq.~(\ref{eq:Pme_int_2}).
In addition, we underline that also the precise determination of $\theta_{13}$ attained 
independently by the reactor experiments Daya Bay and RENO, plays a relevant role
in constraining the new CP-phase $\delta_{14}$, because it helps to constrain the magnitude 
of the leading term in Eq.~(\ref{eq:Pme_atm}) (proportional to $s_{13}^2$) and the amplitude
of the two standard interference terms (which are both proportional to $s_{13}$).
A comparison of our results with those presented in~\cite{palazzo_klop,palazzo_t2k_nova} shows 
that the 68\% and 90\% bounds on $\delta_{14}$ are slightly weaker, despite the 
improved statistics accumulated in the LBL data.
This is due to having taken into account the uncertainty on $\theta_{14}$ and $\theta_{24}$, 
which in~\cite{palazzo_klop,palazzo_t2k_nova} were both fixed to $\sin^2\theta_{14}=\sin^2\theta_{24}=0.025$.

\subsection{Correlation between the two CP-phases\\
 $\delta_{13}$ and $\delta_{14}$}

\begin{figure}[t!]
\vspace*{0.0cm}
\hspace*{-0.15cm}
\includegraphics[width=9.2 cm]{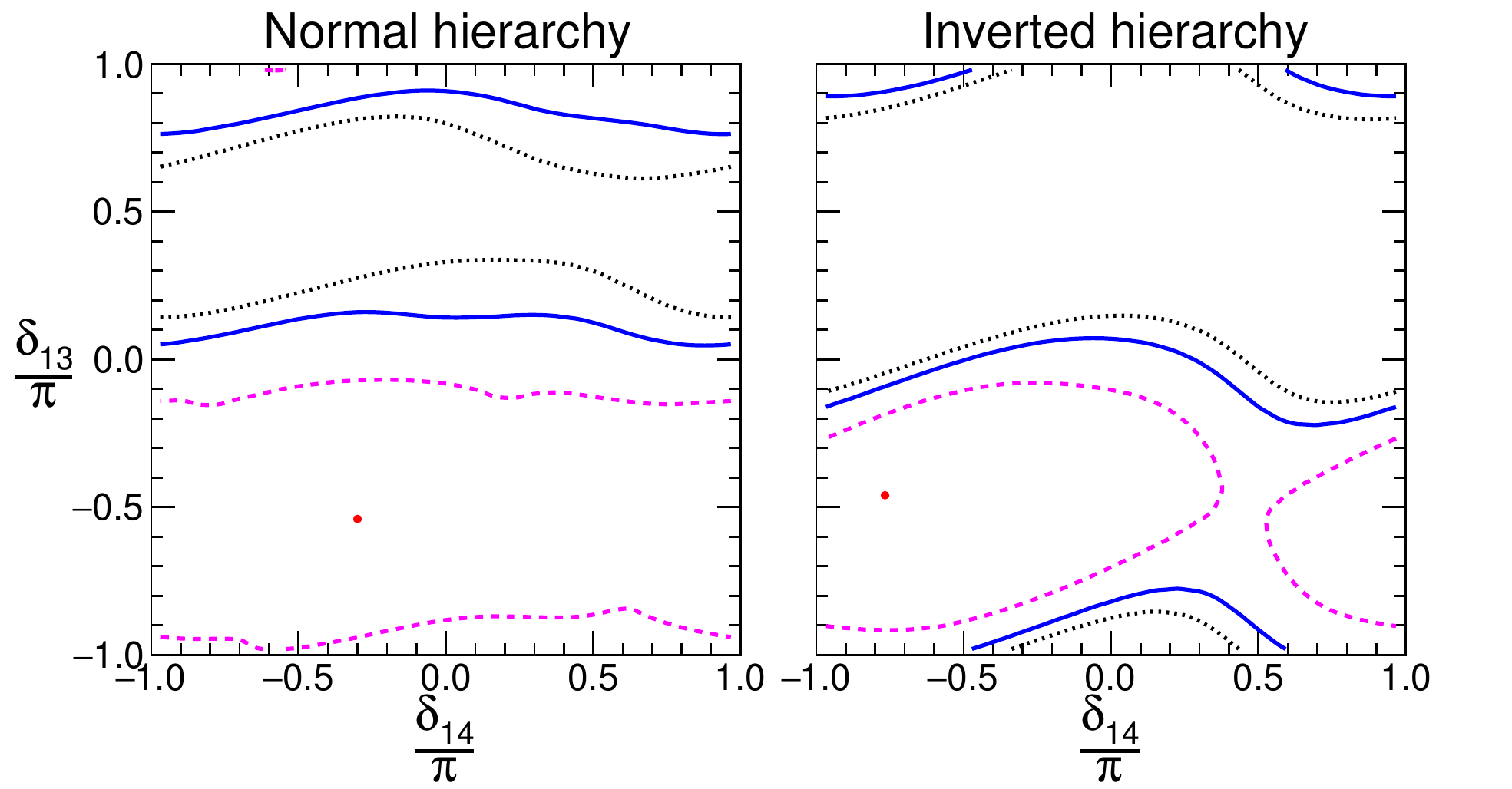}
\vspace*{-0.6cm}
\caption{Regions allowed by the combination of the SBL and LBL data (T2K and NO$\nu$A)
together with the $\theta_{13}$-sensitive reactor results for  NH (left panel)
and IH (right panel) in the plane spanned by the two CP phases $\delta_{13}$ and $\delta_{14}$.
The confidence levels are the same ones used in Fig.~1.
\label{fig:2pan_cp-phases}}
\end{figure}  

Figure~\ref{fig:2pan_cp-phases} shows the constraints  in the plane  of the two CP-phases
[$\delta_{14}$, $\delta_{13}$] for NH (IH) left panel (right panel). Also in this figure, the regions
are obtained combining the SBL data, the LBL results from NO$\nu$A and T2K, and the
data from Daya-Bay and RENO. In both mass hierarchies the CP-conserving cases $\delta_{13}=0, \pi$
are disfavored at $\Delta\chi^2\simeq 2.7$. The best fit value  $\delta_{13}\simeq -\pi/2$,
is basically the same obtained in the 3-flavor case (see the 
analyses~\cite{global_analysis_2016,marrone_neutrino_2016,Esteban:2016qun,Forero:2014bxa}).
This preference comes from the observation of an excess (deficit) of $\nu_e$ ($\bar{\nu}_e$) events with respect to the expectations for the appearance channel $\nu_{\mu}\rightarrow\nu_e$ ($\bar{\nu}_{\mu}\rightarrow\bar{\nu}_e$), when assuming a value of $\theta_{13}$ equal to the best fit point of reactor experiments.
In fact, Eq.~(\ref{eq:Pme_int_1}) shows that, around the first oscillation maximum
($\Delta = \pi/2$), the standard interference term is proportional to $\sin \delta_{13}$.%
\footnote{We recall that the when passing from neutrino to antineutrino probability one has to invert
the sign of all the CP-phases.}
This implies that this term is maximized (minimized) for neutrinos (antineutrinos) for $\delta_{13}=-\pi/2$
in agreement with the observed pattern. Our numerical analysis in the 3+1 scheme reveals that the
presence of the new interference terms does not spoil this picture. This behavior can be explained 
by observing that at the first oscillation maximum  ($\Delta = \pi/2$) the new interference term is
proportional to $\cos (\delta_{13} -\delta_{14})$, and therefore (in contrast to the standard term) 
its sign is the same for neutrinos and antineutrinos. We observe that for $\delta_{13}\simeq \delta_{14}\simeq -\pi/2$, 
the new interference term assumes its maximal positive value (for both neutrinos and antineutrinos). 
In the fit the neutrino dataset dominates over the (lower statistics) antineutrino data set and, 
as a consequence, the excess of $\nu_e$'s wins over the deficit of $\bar \nu_e$'s,
driving the new CP-phase to a best fit value close to $\delta_{14}\simeq -\pi/2$. Finally,
we note that the constraints on the new CP-phase $\delta_{14}$ are very weak. This 
is imputable to the smaller amplitude of the new interference term when compared
to the standard interference term.

\subsection{Impact of sterile neutrinos on the standard neutrino properties}

In the previous subsections we have focused our discussion on the new parameters of the 3+1 scheme
and to the correlation among the two CP-phases. However, it is of interest to see what happens
to the estimates of the standard parameters, which were marginalized in the figures shown until now.
In particular, it seems of particular interest to assess the robustness/fragility of the estimate
of the CP-phase $\delta \equiv \delta_{13}$, the mass hierarchy and the mixing angle $\theta_{23}$, 
which all are at the center of current investigations. 

\begin{figure}[t!]
\vspace*{0cm}
\hspace*{-0.10cm}
\includegraphics[width=9.0 cm]{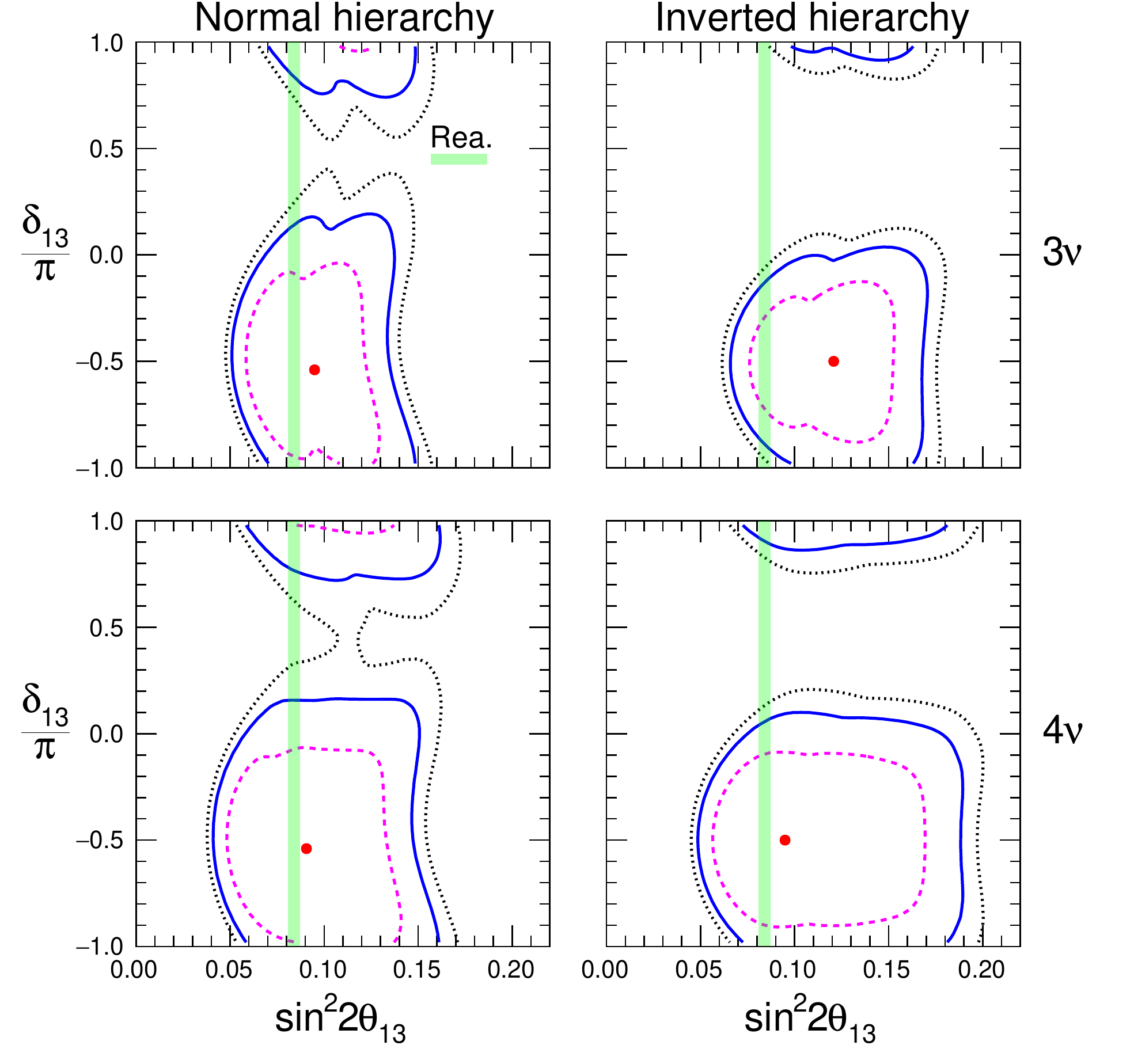}
\vspace*{-0.7cm}
\caption{Regions allowed in the plane [$\sin^2 2\theta_{13}$, $\delta_{13}$] 
 by the joint analysis of all the SBL experiments and the LBL experiments
(T2K and NO$\nu$A). The interval of $\theta_{13}$ identified by the reactor experiments (green vertical band) is displayed for the sake of comparison. The left (right) panels represent the NH (IH) case. 
The upper (lower) panels refer to the 3-flavor (4-flavor) scheme.
The confidence levels are the same reported in Fig.~1.
\label{fig:4pan_3nu_4nu}}
\end{figure}

\begin{figure}[t!]
\vspace*{0cm}
\hspace*{-0.10cm}
\includegraphics[width=9.0 cm]{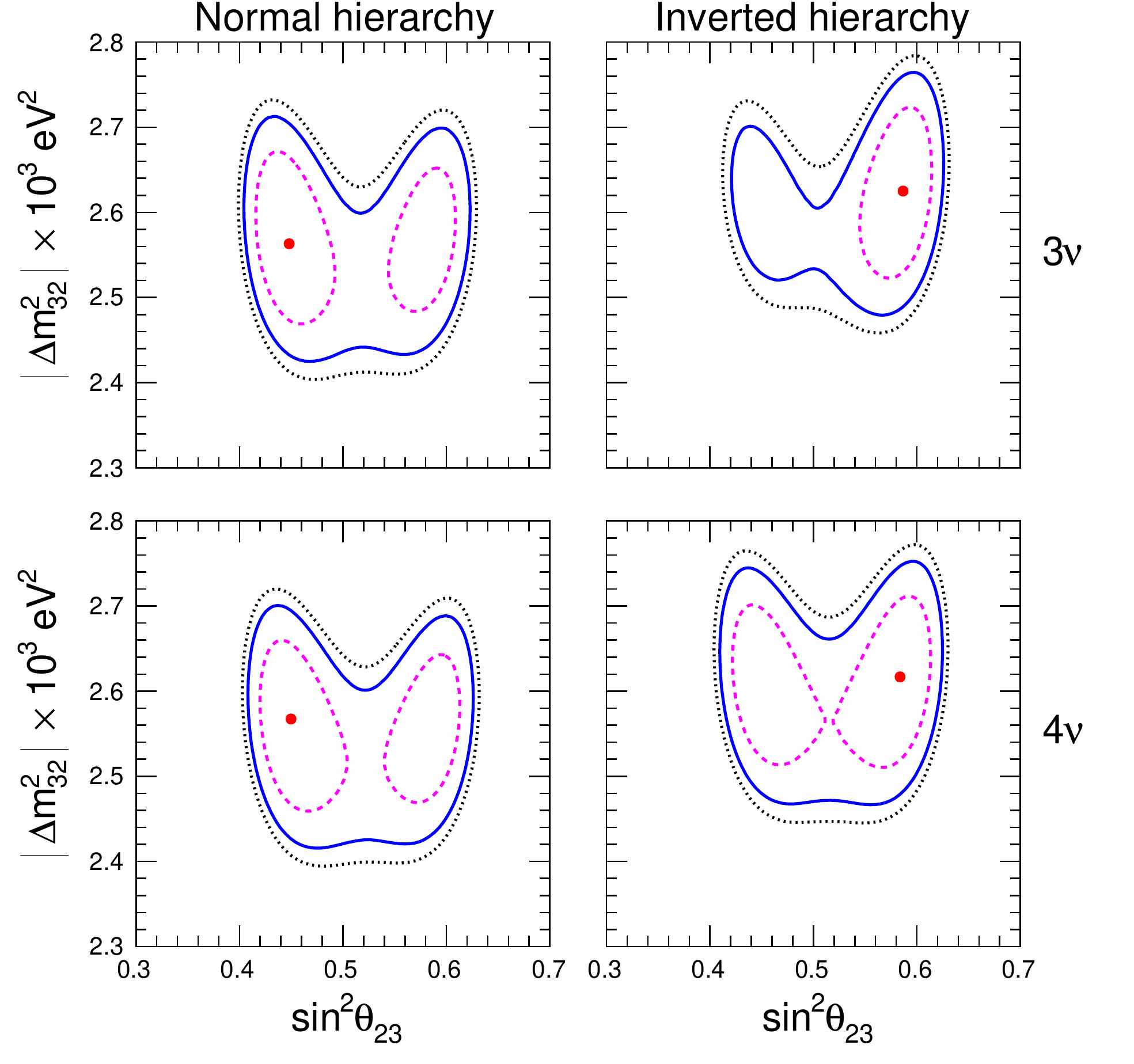}
\vspace*{-0.7cm}
\caption{Regions allowed in the plane [$\sin^2 \theta_{23}$, $\Delta m^2_{32}$] 
 by the joint analysis of all the SBL, the LBL data (T2K and NO$\nu$A),
together with the $\theta_{13}$-sensitive reactor results.  The left (right) panels represent the NH (IH) case. 
The upper (lower) panels refer to the 3-flavor (4-flavor) scheme.
The confidence levels are the same reported in Fig.~1.
\label{fig:4pan_3nu_4nu_octant}}
\end{figure}

Figure~\ref{fig:4pan_3nu_4nu} displays the regions allowed in the plane [$\sin^2 2\theta_{13}, \delta_{13}$] by the joint analysis of all the SBL experiments and the two LBL experiments T2K and NO$\nu$A. 
The two upper panels correspond to the 3-flavor framework,%
\footnote{It should be noted that at the SBL experiments the 3-flavor effects are completely negligible.
Consequently, one can adopt two different approaches when considering the 3-flavor scheme:
i) include the SBL data in the fit,  ii) exclude them from the fit. What changes between
the two approaches is only the value of the absolute minimum of the $\chi^2$. Following the 
first option, one obtains a much higher value than following the second one. This just corresponds 
to the fact that in the 3+1 scheme the goodness of fit increases, because the sterile oscillations
are able to fit the SBL data.  However, when one is interested in parameter estimation, only the
expansion of the $\chi^2$ around its absolute minimum matters and the choice of including
or not including the SBL data in the fit is irrelevant.}
while the two lower ones are obtained in the 4-flavor scheme.
The two left (right) panels refer to NH (IH).  The interval of $\theta_{13}$ identified by the reactor experiments
at 68\% C.L. (represented by the green vertical band) is displayed for the sake of comparison. 
In all cases $\Delta m^2_{32}$ and the mixing angle $\theta_{23}$ are marginalized away. In addition, 
in the 4-flavor case, we marginalize over the two mixing angles ($\theta_{14}$, $\theta_{24}$) 
and the CP-phase $\delta_{14}$. The contours represented in the plots 
correspond to the same confidence levels reported in the previous plots.
The comparison of the 3-flavor and 4-flavor allowed regions shows the following features:
i) the range allowed by LBL alone for $\theta_{13}$ is appreciably larger in the 4-flavor case.
This is a consequence of the presence of the new interference term, which allows
larger excursions of the transition probability from its average value.
However, one can understand that, once the reactor data sensitive to $\theta_{13}$ 
(Daya Bay and RENO) are included in the fit, $\theta_{13}$ is
 ``fixed'' with high precision in both 3-flavor and 4-flavor schemes;
ii) the constraints on the CP-phase $\delta_{13}$ are basically identical in the two schemes.
In both cases there is a preference (rejection) of values of $\sin \delta_{13}<0$ 
($\sin \delta_{13}>0$). We have already discussed this point in the description of
Fig.~\ref{fig:2pan_cp-phases} concerning the correlation on the two CP-phases;
iii) in both schemes the allowed regions, at low confidence levels, present two lobes,
which are more pronounced in the 3-flavor case. This feature is imputable to the 
swap of the best fit value of $\theta_{23}$ among the two quasi-degenerate 
 non-maximal solutions, one in the lower octant (LO) and the other one in the higher octant (HO).
 We will discuss further this point when commenting Fig.~\ref{fig:4pan_3nu_4nu_octant}.

Figure~\ref{fig:4pan_3nu_4nu} also evidences appreciable differences 
between the two cases of NH and IH,  which can be traced to the presence of the matter effects. 
As discussed in Sec.~IV, the matter potential tends to increase (decrease) the theoretically 
expected $\nu_e$ rate in the case of NH (IH). The opposite is true for $\bar \nu_e$'s but their 
weight in the analysis is lower, so the neutrino data sets dominate. In addition, as discussed in
Sec.~IV, the NO$\nu$A $\nu_e$ data are more sensitive than the T2K  $\nu_e$ data 
to the matter effects. More specifically,  the following differences among the two hierarchies emerge,
which are present both in the 3-flavor and 4-flavor schemes. The regions obtained for the
case of IH are shifted towards larger values of $\theta_{13}$ and are slightly wider in the variable $\theta_{13}$ with respect to those obtained in the NH case. In addition,  in the IH  case, the fit 
tends to prefer (reject) values of $\sin \delta_{13} <0$ ($\sin \delta_{13} >0$) in a more pronounced way.

After marginalizing over all parameters we can calculate the $\Delta\chi^2(\text{IH-NH})$
difference between normal and inverted hierarchy
\begin{equation}
\Delta\chi^2(\text{IH-NH}) = \chi^2_{\text{min}}(\text{IH})-\chi^2_{\text{min}}(\text{NH})\ .
\label{delta_chi2_IH_NH}
\end{equation}
For the 3-flavors (4-flavors) analysis of the LBL data alone we obtain $\Delta\chi^2(\text{IH-NH})\simeq 1.0$ (0.8). Therefore this data are (still) not sensitive to the mass hierarchy. The situation sensibly 
changes when the reactor experiments sensitive to $\theta_{13}$ are included in the fit.
In fact, the combination of LBL and reactor provides a slight preference for NH:
$\Delta\chi^2(\text{IH-NH})\simeq 2.0$ (1.3) in the 3-flavor (4-flavors) case. 
The reduced value obtained in the 3+1 framework is due to the inevitable widening of the
parameter space in the presence of an additional neutrino. The preference for the NH case
can be understood comparing the allowed regions from T2K and NO$\nu$A with the constraint on
$\sin^22\theta_{13}$ from reactor experiments (vertical green band in Fig.~\ref{fig:4pan_3nu_4nu}.
One notes that there is a better agreement for NH, whereas for IH the separation between the two
best fit points is at the level of about $\sim 1\sigma$.

Let us now come to the estimate of the standard mixing angle $\theta_{23}$. Recently,
the disappearance analysis of the NO$\nu$A collaboration~\cite{nova_neutrino_2016}
has reported a preference for non-maximal $\theta_{23}$ at the level of 2.5$\sigma$. 
The latest 3-flavor global fits~\cite{marrone_neutrino_2016, Esteban:2016qun} have shown
that this feature persists at the level of about 2$\sigma$ even when other datasets are included in the analysis.
Given the important role of the atmospheric angle $\theta_{23}$ in the context of model
building, it seems opportune to assess the estimate of such a parameter in the enlarged 3+1 scheme.  

We recall that, in the 3-flavor framework,  the disappearance channel is sensitive
to possible deviations from maximal mixing but it is blind to the octant of $\theta_{23}$.
This occurs because the $\nu_\mu \to \nu_\mu$ disappearance probability is proportional
to $\sin^2 2\theta_{23}$. Therefore, if only the disappearance channel data 
are included in the analysis, the allowed ranges are symmetrical with respect to $\sin^2 \theta_{23} = 0.5$.
This symmetry is broken when one considers also the appearance channel. 
This happens because the $\nu_\mu \to \nu_e$ transition probability is octant sensitive 
since its leading term depends on $\sin^2 \theta_{23}$. In the 4-flavor scheme, the disappearance probability
remains basically unaltered, so one expects that the sensitivity to potential 
deviations from maximal mixing remains unaltered. In contrast, the appearance probability
is profoundly affected by the new interference term, which, as recently shown in~\cite{Agarwalla:2016xlg},
leads to a loss of sensitivity to the octant of $\theta_{23}$. 

Figure~\ref{fig:4pan_3nu_4nu_octant} reports the allowed regions in the plane 
$[\sin^{2}\theta_{23},\Delta m^{2}_{32}]$, all the other parameters having 
being marginalized away. The left (right) panels refer to normal (inverted) hierarchy, while
the upper (lower) panels refer to the 3-flavor (4-flavor) case.
In both schemes we have included in the analysis all the SBL data,  
the LBL results from T2K and NO$\nu$A (both appearance and disappearance channels)
and the $\theta_{13}$-sensitive reactor experiments. 
The results reported in the upper panels show a weak preference for non-maximal mixing in the
3-flavor scenario. 
We note that there is a change in the preferred octant when switching from normal to inverted
hierarchy. This is a consequence of the anticorrelation between $\theta_{13}$ and
$\theta_{23}$, introduced by the appearance data set: the lower $\theta_{13}$ the higher 
the value of $\theta_{23}$. In NH we find a negligible preference for the lower
octant ($\theta_{23}<45^0$). In IH the effect is more pronounced and the higher 
octant ($\theta_{23}>45^0$) is favored at a non-negligible statistical level.
The two lower panels depict how the situation changes in the 4-flavor scheme. 
We can observe that the allowed regions becomes basically symmetric
around maximal mixing. As expected from the discussion above, in the 4-flavor scheme, 
the sensitivity to the $\theta_{23}$ octant gets lost.

\begin{figure}[t!]
\vspace*{0cm}
\hspace*{-0.10cm}
\includegraphics[width=9.0 cm]{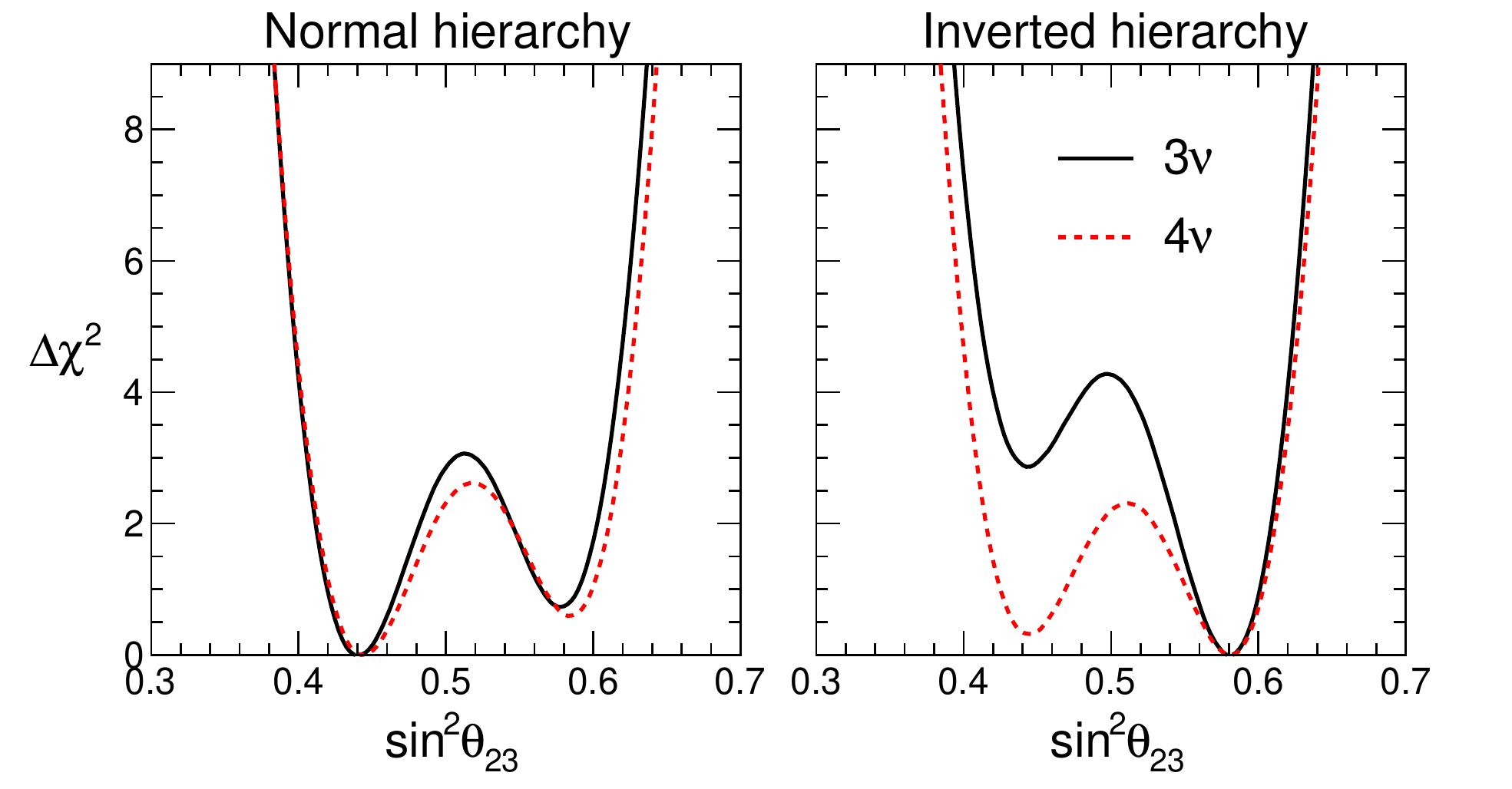}
\vspace*{-0.7cm}
\caption{Marginalized $\Delta \chi^2$ for the parameter $\sin^2 \theta_{23}$
for NH (left panel) and IH (right panel). The black solid line indicates the 3-flavor 
case, while the red dashed line refers to the 3+1 scheme. 
\label{fig:theta_23}}
\end{figure}  
 
In order to clarify this picture, we present in Fig.~6 the marginalized 
$\Delta \chi^2$ as a function of $\sin^2 \theta_{23}$.
The left (right) panel corresponds to NH (IH). 
The black solid line indicates the 3-flavor case, while the red dashed line 
refers to the 3+1 scheme. In both NH and IH cases,
in the 4-flavor scheme non-maximal mixing is disfavored approximately
at $\Delta \chi^2 \simeq 2.5$ (corresponding almost to 90\% C.L. for 1 d.o.f.). 
Therefore, the weak preference for non-maximal $\theta_{23}$ originating from (part of) the disappearance channel data
is a stable feature, which is independent of the scheme adopted (3-flavor or 4-flavor). In contrast, we see that the preference
for $\theta_{23}>45^0$ found in IH completely disappears in the 3+1 scheme. 
We finally note that this behavior is in line with the results of the sensitivity study performed
in~\cite{Agarwalla:2016xlg}, where it has been shown that even in a future experiment like DUNE, 
which will make use of a high-intensity broad-band neutrino beam, the sensitivity to the octant 
drastically decreases in the 3+1 scheme. Our analysis performed with the real data confirms 
such a general behavior, showing that the indication on the octant of $\theta_{23}$ becomes 
a fragile feature in the 3+1 framework.\\

\section{Conclusions}

 We have shown that, within the 3+1 scheme,  the combination of the existing SBL data with the LBL results coming from the two currently running experiments NO$\nu$A and T2K, enables us to simultaneously constrain two active-sterile mixing angles $\theta_{14}$ and $\theta_{24}$ and two CP-phases $\delta_{13} \equiv \delta$ and $\delta_{14}$, although the constraints on this
 last CP-phase are still weak. The two mixing angles are basically determined by the SBL data, while the two CP-phases are identified by the LBL experiments, once the information coming from the SBL setups is taken into account. We have also assessed the robustness/fragility of the estimates of the standard 3-flavor properties in the more general 3+1 scheme. To this regard we found that: i) the indication of CP-violation found in the 3-flavor analyses persists also in the 3+1 scheme, with $\delta_{13} \equiv \delta$ having still its best fit value around $-\pi/2$; ii) the 3-flavor weak hint in favor of the normal hierarchy becomes even less significant when sterile neutrinos come into play; iii) the weak indication of non-maximal $\theta_{23}$  (driven by NO$\nu$A disappearance data) persists in the 3+1 scheme, where maximal mixing is disfavored at almost the 90\% C.L. in both normal and inverted mass hierarchy; iv) the preference in favor of one of the two octants of $\theta_{23}$ found in the 3-flavor framework (higher octant for inverted mass hierarchy) is completely washed out in the 3+1 scheme. We hope that our joint analysis of SBL and LBL data in the 3+1 scheme may serve as a guide for more comprehensive analyses and may increase the awareness of the neutrino community towards the important role of LBL experiments in the search of CP violation induced by light sterile neutrinos.

\section*{Acknowledgments}

This work was partially supported by the research grant {\sl Theoretical Astroparticle Physics} number 2012CPPYP7 under the program PRIN 2012 funded by the Italian Ministero dell'Istruzione, Universit\`a e della Ricerca (MIUR) and by the research project 
{\em TAsP} funded by the Instituto Nazionale di Fisica Nucleare (INFN). A. P. is supported  by the project 
{\em Beyond three neutrino families} within the FutureInResearch program, Fondo di Sviluppo e Coesione 2007-2013, APQ Ricerca Regione Puglia ÒProgramma regionale a sostegno della specializzazione intelligente e della  sostenibilit\`a sociale ed ambientale.

\bibliography{lbl_sbl_arxiv_v3} 

\begin{thebibliography}{82}
\expandafter\ifx\csname natexlab\endcsname\relax\def\natexlab#1{#1}\fi
\expandafter\ifx\csname bibnamefont\endcsname\relax
  \def\bibnamefont#1{#1}\fi
\expandafter\ifx\csname bibfnamefont\endcsname\relax
  \def\bibfnamefont#1{#1}\fi
\expandafter\ifx\csname citenamefont\endcsname\relax
  \def\citenamefont#1{#1}\fi
\expandafter\ifx\csname url\endcsname\relax
  \def\url#1{\texttt{#1}}\fi
\expandafter\ifx\csname urlprefix\endcsname\relax\def\urlprefix{URL }\fi
\providecommand{\bibinfo}[2]{#2}
\providecommand{\eprint}[2][]{\url{#2}}

\bibitem[{\citenamefont{Klop and Palazzo}(2015)}]{palazzo_klop}
\bibinfo{author}{\bibfnamefont{N.}~\bibnamefont{Klop}} \bibnamefont{and}
  \bibinfo{author}{\bibfnamefont{A.}~\bibnamefont{Palazzo}},
  \bibinfo{journal}{Phys. Rev.} \textbf{\bibinfo{volume}{D91}},
  \bibinfo{pages}{073017} (\bibinfo{year}{2015}), \eprint{1412.7524}.

\bibitem[{\citenamefont{Palazzo}(2016)}]{palazzo_t2k_nova}
\bibinfo{author}{\bibfnamefont{A.}~\bibnamefont{Palazzo}},
  \bibinfo{journal}{Phys. Lett.} \textbf{\bibinfo{volume}{B757}},
  \bibinfo{pages}{142} (\bibinfo{year}{2016}), \eprint{1509.03148}.

\bibitem[{\citenamefont{Capozzi et~al.}(2016)\citenamefont{Capozzi, Lisi,
  Marrone, Montanino, and Palazzo}}]{global_analysis_2016}
\bibinfo{author}{\bibfnamefont{F.}~\bibnamefont{Capozzi}},
  \bibinfo{author}{\bibfnamefont{E.}~\bibnamefont{Lisi}},
  \bibinfo{author}{\bibfnamefont{A.}~\bibnamefont{Marrone}},
  \bibinfo{author}{\bibfnamefont{D.}~\bibnamefont{Montanino}},
  \bibnamefont{and} \bibinfo{author}{\bibfnamefont{A.}~\bibnamefont{Palazzo}},
  \bibinfo{journal}{Nucl. Phys.} \textbf{\bibinfo{volume}{B908}},
  \bibinfo{pages}{218} (\bibinfo{year}{2016}), \eprint{1601.07777}.

\bibitem[{\citenamefont{Marrone}(2016)}]{marrone_neutrino_2016}
\bibinfo{author}{\bibfnamefont{A.}~\bibnamefont{Marrone}}
  (\bibinfo{year}{2016}), \bibinfo{note}{presented at the 27th International
  Conference on Neutrino Physics and Astrophysics in July 2016 in London,}.

\bibitem[{\citenamefont{Esteban et~al.}(2016)\citenamefont{Esteban,
  Gonzalez-Garcia, Maltoni, Martinez-Soler, and Schwetz}}]{Esteban:2016qun}
\bibinfo{author}{\bibfnamefont{I.}~\bibnamefont{Esteban}},
  \bibinfo{author}{\bibfnamefont{M.~C.} \bibnamefont{Gonzalez-Garcia}},
  \bibinfo{author}{\bibfnamefont{M.}~\bibnamefont{Maltoni}},
  \bibinfo{author}{\bibfnamefont{I.}~\bibnamefont{Martinez-Soler}},
  \bibnamefont{and} \bibinfo{author}{\bibfnamefont{T.}~\bibnamefont{Schwetz}}
  (\bibinfo{year}{2016}), \eprint{1611.01514}.

\bibitem[{\citenamefont{Forero et~al.}(2014)\citenamefont{Forero, Tortola, and
  Valle}}]{Forero:2014bxa}
\bibinfo{author}{\bibfnamefont{D.~V.} \bibnamefont{Forero}},
  \bibinfo{author}{\bibfnamefont{M.}~\bibnamefont{Tortola}}, \bibnamefont{and}
  \bibinfo{author}{\bibfnamefont{J.~W.~F.} \bibnamefont{Valle}},
  \bibinfo{journal}{Phys.Rev.} \textbf{\bibinfo{volume}{D90}},
  \bibinfo{pages}{093006} (\bibinfo{year}{2014}), \eprint{arXiv:1405.7540}.

\bibitem[{\citenamefont{Vahle}(2016)}]{nova_neutrino_2016}
\bibinfo{author}{\bibfnamefont{P.}~\bibnamefont{Vahle}} (\bibinfo{year}{2016}),
  \bibinfo{note}{presented at the 27th International Conference on Neutrino
  Physics and Astrophysics in July 2016 in London,}.

\bibitem[{\citenamefont{Iwamoto}(2016)}]{t2k_ichep_2016}
\bibinfo{author}{\bibfnamefont{K.}~\bibnamefont{Iwamoto}}
  (\bibinfo{year}{2016}), \bibinfo{note}{presented at the 38th International
  Conference on High Energy Physics in August 2016 in Chicago,}.

\bibitem[{\citenamefont{Giunti et~al.}(2013)\citenamefont{Giunti, Laveder, Li,
  and Long}}]{Giunti:2013aea}
\bibinfo{author}{\bibfnamefont{C.}~\bibnamefont{Giunti}},
  \bibinfo{author}{\bibfnamefont{M.}~\bibnamefont{Laveder}},
  \bibinfo{author}{\bibfnamefont{Y.}~\bibnamefont{Li}}, \bibnamefont{and}
  \bibinfo{author}{\bibfnamefont{H.}~\bibnamefont{Long}},
  \bibinfo{journal}{Phys.Rev.} \textbf{\bibinfo{volume}{D88}},
  \bibinfo{pages}{073008} (\bibinfo{year}{2013}), \eprint{arXiv:1308.5288}.

\bibitem[{\citenamefont{Kopp et~al.}(2013)\citenamefont{Kopp, Machado, Maltoni,
  and Schwetz}}]{kopp_sterile}
\bibinfo{author}{\bibfnamefont{J.}~\bibnamefont{Kopp}},
  \bibinfo{author}{\bibfnamefont{P.~A.~N.} \bibnamefont{Machado}},
  \bibinfo{author}{\bibfnamefont{M.}~\bibnamefont{Maltoni}}, \bibnamefont{and}
  \bibinfo{author}{\bibfnamefont{T.}~\bibnamefont{Schwetz}},
  \bibinfo{journal}{JHEP} \textbf{\bibinfo{volume}{05}}, \bibinfo{pages}{050}
  (\bibinfo{year}{2013}), \eprint{1303.3011}.

\bibitem[{\citenamefont{Collin et~al.}(2016{\natexlab{a}})\citenamefont{Collin,
  Arguelles, Conrad, and Shaevitz}}]{Collin:2016rao}
\bibinfo{author}{\bibfnamefont{G.~H.} \bibnamefont{Collin}},
  \bibinfo{author}{\bibfnamefont{C.~A.} \bibnamefont{Arguelles}},
  \bibinfo{author}{\bibfnamefont{J.~M.} \bibnamefont{Conrad}},
  \bibnamefont{and} \bibinfo{author}{\bibfnamefont{M.~H.}
  \bibnamefont{Shaevitz}}, \bibinfo{journal}{Nucl.Phys.}
  \textbf{\bibinfo{volume}{B908}}, \bibinfo{pages}{354}
  (\bibinfo{year}{2016}{\natexlab{a}}), \eprint{arXiv:1602.00671}.

\bibitem[{\citenamefont{Collin et~al.}(2016{\natexlab{b}})\citenamefont{Collin,
  ArgŸelles, Conrad, and Shaevitz}}]{Collin:2016aqd}
\bibinfo{author}{\bibfnamefont{G.~H.} \bibnamefont{Collin}},
  \bibinfo{author}{\bibfnamefont{C.~A.} \bibnamefont{ArgŸelles}},
  \bibinfo{author}{\bibfnamefont{J.~M.} \bibnamefont{Conrad}},
  \bibnamefont{and} \bibinfo{author}{\bibfnamefont{M.~H.}
  \bibnamefont{Shaevitz}}, \bibinfo{journal}{Phys. Rev. Lett.}
  \textbf{\bibinfo{volume}{117}}, \bibinfo{pages}{221801}
  (\bibinfo{year}{2016}{\natexlab{b}}), \eprint{1607.00011}.

\bibitem[{\citenamefont{Agarwalla
  et~al.}(2016{\natexlab{a}})\citenamefont{Agarwalla, Chatterjee, Dasgupta, and
  Palazzo}}]{palazzo_t2k_nova_sterile}
\bibinfo{author}{\bibfnamefont{S.~K.} \bibnamefont{Agarwalla}},
  \bibinfo{author}{\bibfnamefont{S.~S.} \bibnamefont{Chatterjee}},
  \bibinfo{author}{\bibfnamefont{A.}~\bibnamefont{Dasgupta}}, \bibnamefont{and}
  \bibinfo{author}{\bibfnamefont{A.}~\bibnamefont{Palazzo}},
  \bibinfo{journal}{JHEP} \textbf{\bibinfo{volume}{02}}, \bibinfo{pages}{111}
  (\bibinfo{year}{2016}{\natexlab{a}}), \eprint{1601.05995}.

\bibitem[{\citenamefont{Agarwalla
  et~al.}(2016{\natexlab{b}})\citenamefont{Agarwalla, Chatterjee, and
  Palazzo}}]{palazzo_dune_sterile}
\bibinfo{author}{\bibfnamefont{S.~K.} \bibnamefont{Agarwalla}},
  \bibinfo{author}{\bibfnamefont{S.~S.} \bibnamefont{Chatterjee}},
  \bibnamefont{and} \bibinfo{author}{\bibfnamefont{A.}~\bibnamefont{Palazzo}},
  \bibinfo{journal}{JHEP} \textbf{\bibinfo{volume}{09}}, \bibinfo{pages}{016}
  (\bibinfo{year}{2016}{\natexlab{b}}), \eprint{1603.03759}.

\bibitem[{\citenamefont{Bilenky et~al.}(1998)\citenamefont{Bilenky, Giunti, and
  Grimus}}]{Bilenky:1996rw}
\bibinfo{author}{\bibfnamefont{S.~M.} \bibnamefont{Bilenky}},
  \bibinfo{author}{\bibfnamefont{C.}~\bibnamefont{Giunti}}, \bibnamefont{and}
  \bibinfo{author}{\bibfnamefont{W.}~\bibnamefont{Grimus}},
  \bibinfo{journal}{Eur. Phys. J.} \textbf{\bibinfo{volume}{C1}},
  \bibinfo{pages}{247} (\bibinfo{year}{1998}), \eprint{hep-ph/9607372}.

\bibitem[{\citenamefont{Athanassopoulos et~al.}(1995)}]{Athanassopoulos:1995iw}
\bibinfo{author}{\bibfnamefont{C.}~\bibnamefont{Athanassopoulos}}
  \bibnamefont{et~al.} (\bibinfo{collaboration}{LSND}), \bibinfo{journal}{Phys.
  Rev. Lett.} \textbf{\bibinfo{volume}{75}}, \bibinfo{pages}{2650}
  (\bibinfo{year}{1995}), \eprint{nucl-ex/9504002}.

\bibitem[{\citenamefont{Aguilar et~al.}(2001)}]{Aguilar:2001ty}
\bibinfo{author}{\bibfnamefont{A.}~\bibnamefont{Aguilar}} \bibnamefont{et~al.}
  (\bibinfo{collaboration}{LSND}), \bibinfo{journal}{Phys. Rev.}
  \textbf{\bibinfo{volume}{D64}}, \bibinfo{pages}{112007}
  (\bibinfo{year}{2001}), \eprint{hep-ex/0104049}.

\bibitem[{\citenamefont{Abdurashitov et~al.}(2006)}]{Abdurashitov:2005tb}
\bibinfo{author}{\bibfnamefont{J.~N.} \bibnamefont{Abdurashitov}}
  \bibnamefont{et~al.} (\bibinfo{collaboration}{SAGE}), \bibinfo{journal}{Phys.
  Rev.} \textbf{\bibinfo{volume}{C73}}, \bibinfo{pages}{045805}
  (\bibinfo{year}{2006}), \eprint{nucl-ex/0512041}.

\bibitem[{\citenamefont{Laveder}(2007)}]{Laveder:2007zz}
\bibinfo{author}{\bibfnamefont{M.}~\bibnamefont{Laveder}},
  \bibinfo{journal}{Nucl. Phys. Proc. Suppl.} \textbf{\bibinfo{volume}{168}},
  \bibinfo{pages}{344} (\bibinfo{year}{2007}).

\bibitem[{\citenamefont{Giunti and Laveder}(2007)}]{Giunti:2006bj}
\bibinfo{author}{\bibfnamefont{C.}~\bibnamefont{Giunti}} \bibnamefont{and}
  \bibinfo{author}{\bibfnamefont{M.}~\bibnamefont{Laveder}},
  \bibinfo{journal}{Mod. Phys. Lett.} \textbf{\bibinfo{volume}{A22}},
  \bibinfo{pages}{2499} (\bibinfo{year}{2007}), \eprint{hep-ph/0610352}.

\bibitem[{\citenamefont{Giunti and
  Laveder}(2011{\natexlab{a}})}]{Giunti:2010zu}
\bibinfo{author}{\bibfnamefont{C.}~\bibnamefont{Giunti}} \bibnamefont{and}
  \bibinfo{author}{\bibfnamefont{M.}~\bibnamefont{Laveder}},
  \bibinfo{journal}{Phys. Rev.} \textbf{\bibinfo{volume}{C83}},
  \bibinfo{pages}{065504} (\bibinfo{year}{2011}{\natexlab{a}}),
  \eprint{arXiv:1006.3244}.

\bibitem[{\citenamefont{Giunti et~al.}(2012)\citenamefont{Giunti, Laveder, Li,
  Liu, and Long}}]{Giunti:2012tn}
\bibinfo{author}{\bibfnamefont{C.}~\bibnamefont{Giunti}},
  \bibinfo{author}{\bibfnamefont{M.}~\bibnamefont{Laveder}},
  \bibinfo{author}{\bibfnamefont{Y.}~\bibnamefont{Li}},
  \bibinfo{author}{\bibfnamefont{Q.}~\bibnamefont{Liu}}, \bibnamefont{and}
  \bibinfo{author}{\bibfnamefont{H.}~\bibnamefont{Long}},
  \bibinfo{journal}{Phys. Rev.} \textbf{\bibinfo{volume}{D86}},
  \bibinfo{pages}{113014} (\bibinfo{year}{2012}), \eprint{arXiv:1210.5715}.

\bibitem[{\citenamefont{Kaether et~al.}(2010)\citenamefont{Kaether, Hampel,
  Heusser, Kiko, and Kirsten}}]{Kaether:2010ag}
\bibinfo{author}{\bibfnamefont{F.}~\bibnamefont{Kaether}},
  \bibinfo{author}{\bibfnamefont{W.}~\bibnamefont{Hampel}},
  \bibinfo{author}{\bibfnamefont{G.}~\bibnamefont{Heusser}},
  \bibinfo{author}{\bibfnamefont{J.}~\bibnamefont{Kiko}}, \bibnamefont{and}
  \bibinfo{author}{\bibfnamefont{T.}~\bibnamefont{Kirsten}},
  \bibinfo{journal}{Phys. Lett.} \textbf{\bibinfo{volume}{B685}},
  \bibinfo{pages}{47} (\bibinfo{year}{2010}), \eprint{arXiv:1001.2731}.

\bibitem[{\citenamefont{Abdurashitov et~al.}(2009)}]{Abdurashitov:2009tn}
\bibinfo{author}{\bibfnamefont{J.~N.} \bibnamefont{Abdurashitov}}
  \bibnamefont{et~al.} (\bibinfo{collaboration}{SAGE}), \bibinfo{journal}{Phys.
  Rev.} \textbf{\bibinfo{volume}{C80}}, \bibinfo{pages}{015807}
  (\bibinfo{year}{2009}), \eprint{arXiv:0901.2200}.

\bibitem[{\citenamefont{Mention et~al.}(2011)}]{Mention:2011rk}
\bibinfo{author}{\bibfnamefont{G.}~\bibnamefont{Mention}} \bibnamefont{et~al.},
  \bibinfo{journal}{Phys. Rev.} \textbf{\bibinfo{volume}{D83}},
  \bibinfo{pages}{073006} (\bibinfo{year}{2011}), \eprint{arXiv:1101.2755}.

\bibitem[{\citenamefont{Mueller et~al.}(2011)}]{Mueller:2011nm}
\bibinfo{author}{\bibfnamefont{T.~A.} \bibnamefont{Mueller}}
  \bibnamefont{et~al.}, \bibinfo{journal}{Phys. Rev.}
  \textbf{\bibinfo{volume}{C83}}, \bibinfo{pages}{054615}
  (\bibinfo{year}{2011}), \eprint{arXiv:1101.2663}.

\bibitem[{\citenamefont{Huber}(2011)}]{Huber:2011wv}
\bibinfo{author}{\bibfnamefont{P.}~\bibnamefont{Huber}},
  \bibinfo{journal}{Phys. Rev.} \textbf{\bibinfo{volume}{C84}},
  \bibinfo{pages}{024617} (\bibinfo{year}{2011}), \eprint{arXiv:1106.0687}.

\bibitem[{\citenamefont{Gariazzo et~al.}(2016)\citenamefont{Gariazzo, Giunti,
  Laveder, Li, and Zavanin}}]{Gariazzo:2015rra}
\bibinfo{author}{\bibfnamefont{S.}~\bibnamefont{Gariazzo}},
  \bibinfo{author}{\bibfnamefont{C.}~\bibnamefont{Giunti}},
  \bibinfo{author}{\bibfnamefont{M.}~\bibnamefont{Laveder}},
  \bibinfo{author}{\bibfnamefont{Y.}~\bibnamefont{Li}}, \bibnamefont{and}
  \bibinfo{author}{\bibfnamefont{E.}~\bibnamefont{Zavanin}},
  \bibinfo{journal}{J. Phys.} \textbf{\bibinfo{volume}{G43}},
  \bibinfo{pages}{033001} (\bibinfo{year}{2016}), \eprint{arXiv:1507.08204}.

\bibitem[{\citenamefont{Giunti}(2016)}]{Giunti:2016oan}
\bibinfo{author}{\bibfnamefont{C.}~\bibnamefont{Giunti}}
  (\bibinfo{year}{2016}), \bibinfo{note}{{Proceedings of the 27th International
  Conference on Neutrino Physics and Astrophysics (Neutrino 2016) London,
  United Kingdom, July 4-9, 2016}}, \eprint{1609.04688}.

\bibitem[{\citenamefont{Aguilar-Arevalo et~al.}(2009)}]{AguilarArevalo:2008rc}
\bibinfo{author}{\bibfnamefont{A.~A.} \bibnamefont{Aguilar-Arevalo}}
  \bibnamefont{et~al.} (\bibinfo{collaboration}{MiniBooNE}),
  \bibinfo{journal}{Phys. Rev. Lett.} \textbf{\bibinfo{volume}{102}},
  \bibinfo{pages}{101802} (\bibinfo{year}{2009}), \eprint{arXiv:0812.2243}.

\bibitem[{\citenamefont{Aguilar-Arevalo
  et~al.}(2013)}]{Aguilar-Arevalo:2013pmq}
\bibinfo{author}{\bibfnamefont{A.}~\bibnamefont{Aguilar-Arevalo}}
  \bibnamefont{et~al.} (\bibinfo{collaboration}{MiniBooNE}),
  \bibinfo{journal}{Phys.Rev.Lett.} \textbf{\bibinfo{volume}{110}},
  \bibinfo{pages}{161801} (\bibinfo{year}{2013}), \eprint{arXiv:1303.2588}.

\bibitem[{\citenamefont{Borodovsky et~al.}(1992)}]{Borodovsky:1992pn}
\bibinfo{author}{\bibfnamefont{L.}~\bibnamefont{Borodovsky}}
  \bibnamefont{et~al.} (\bibinfo{collaboration}{BNL-E776}),
  \bibinfo{journal}{Phys. Rev. Lett.} \textbf{\bibinfo{volume}{68}},
  \bibinfo{pages}{274} (\bibinfo{year}{1992}).

\bibitem[{\citenamefont{Armbruster et~al.}(2002)}]{Armbruster:2002mp}
\bibinfo{author}{\bibfnamefont{B.}~\bibnamefont{Armbruster}}
  \bibnamefont{et~al.} (\bibinfo{collaboration}{KARMEN}),
  \bibinfo{journal}{Phys. Rev.} \textbf{\bibinfo{volume}{D65}},
  \bibinfo{pages}{112001} (\bibinfo{year}{2002}), \eprint{hep-ex/0203021}.

\bibitem[{\citenamefont{Astier et~al.}(2003)}]{Astier:2003gs}
\bibinfo{author}{\bibfnamefont{P.}~\bibnamefont{Astier}} \bibnamefont{et~al.}
  (\bibinfo{collaboration}{NOMAD}), \bibinfo{journal}{Phys. Lett.}
  \textbf{\bibinfo{volume}{B570}}, \bibinfo{pages}{19} (\bibinfo{year}{2003}),
  \eprint{hep-ex/0306037}.

\bibitem[{\citenamefont{Antonello et~al.}(2013)}]{Antonello:2013gut}
\bibinfo{author}{\bibfnamefont{M.}~\bibnamefont{Antonello}}
  \bibnamefont{et~al.} (\bibinfo{collaboration}{ICARUS}),
  \bibinfo{journal}{Eur.Phys.J.} \textbf{\bibinfo{volume}{C73}},
  \bibinfo{pages}{2599} (\bibinfo{year}{2013}), \eprint{arXiv:1307.4699}.

\bibitem[{\citenamefont{Agafonova et~al.}(2013)}]{Agafonova:2013xsk}
\bibinfo{author}{\bibfnamefont{N.}~\bibnamefont{Agafonova}}
  \bibnamefont{et~al.} (\bibinfo{collaboration}{OPERA}),
  \bibinfo{journal}{JHEP} \textbf{\bibinfo{volume}{1307}}, \bibinfo{pages}{004}
  (\bibinfo{year}{2013}), \eprint{arXiv:1303.3953}.

\bibitem[{\citenamefont{Palazzo}(2015)}]{Palazzo:2015wea}
\bibinfo{author}{\bibfnamefont{A.}~\bibnamefont{Palazzo}},
  \bibinfo{journal}{Phys.Rev.} \textbf{\bibinfo{volume}{D91}},
  \bibinfo{pages}{091301} (\bibinfo{year}{2015}), \eprint{arXiv:1503.03966}.

\bibitem[{\citenamefont{Declais et~al.}(1994)}]{Declais:1994ma}
\bibinfo{author}{\bibfnamefont{Y.}~\bibnamefont{Declais}} \bibnamefont{et~al.}
  (\bibinfo{collaboration}{Bugey}), \bibinfo{journal}{Phys. Lett.}
  \textbf{\bibinfo{volume}{B338}}, \bibinfo{pages}{383} (\bibinfo{year}{1994}).

\bibitem[{\citenamefont{Kuvshinnikov et~al.}(1991)\citenamefont{Kuvshinnikov,
  Mikaelyan, Nikolaev, Skorokhvatov, and Etenko}}]{Kuvshinnikov:1990ry}
\bibinfo{author}{\bibfnamefont{A.}~\bibnamefont{Kuvshinnikov}},
  \bibinfo{author}{\bibfnamefont{L.}~\bibnamefont{Mikaelyan}},
  \bibinfo{author}{\bibfnamefont{S.}~\bibnamefont{Nikolaev}},
  \bibinfo{author}{\bibfnamefont{M.}~\bibnamefont{Skorokhvatov}},
  \bibnamefont{and} \bibinfo{author}{\bibfnamefont{A.}~\bibnamefont{Etenko}},
  \bibinfo{journal}{JETP Lett.} \textbf{\bibinfo{volume}{54}},
  \bibinfo{pages}{253} (\bibinfo{year}{1991}).

\bibitem[{\citenamefont{Achkar et~al.}(1995)}]{Declais:1995su}
\bibinfo{author}{\bibfnamefont{B.}~\bibnamefont{Achkar}} \bibnamefont{et~al.}
  (\bibinfo{collaboration}{Bugey}), \bibinfo{journal}{Nucl. Phys.}
  \textbf{\bibinfo{volume}{B434}}, \bibinfo{pages}{503} (\bibinfo{year}{1995}).

\bibitem[{\citenamefont{Zacek et~al.}(1986)}]{Zacek:1986cu}
\bibinfo{author}{\bibfnamefont{G.}~\bibnamefont{Zacek}} \bibnamefont{et~al.}
  (\bibinfo{collaboration}{CalTech-SIN-TUM}), \bibinfo{journal}{Phys. Rev.}
  \textbf{\bibinfo{volume}{D34}}, \bibinfo{pages}{2621} (\bibinfo{year}{1986}).

\bibitem[{\citenamefont{Hoummada et~al.}(1995)\citenamefont{Hoummada,
  Lazrak~Mikou, Bagieu, Cavaignac, and Holm~Koang}}]{Hoummada:1995zz}
\bibinfo{author}{\bibfnamefont{A.}~\bibnamefont{Hoummada}},
  \bibinfo{author}{\bibfnamefont{S.}~\bibnamefont{Lazrak~Mikou}},
  \bibinfo{author}{\bibfnamefont{G.}~\bibnamefont{Bagieu}},
  \bibinfo{author}{\bibfnamefont{J.}~\bibnamefont{Cavaignac}},
  \bibnamefont{and}
  \bibinfo{author}{\bibfnamefont{D.}~\bibnamefont{Holm~Koang}},
  \bibinfo{journal}{Applied Radiation and Isotopes}
  \textbf{\bibinfo{volume}{46}}, \bibinfo{pages}{449} (\bibinfo{year}{1995}).

\bibitem[{\citenamefont{Vidyakin et~al.}(1990)}]{Vidyakin:1990iz}
\bibinfo{author}{\bibfnamefont{G.~S.} \bibnamefont{Vidyakin}}
  \bibnamefont{et~al.} (\bibinfo{collaboration}{Krasnoyarsk}),
  \bibinfo{journal}{Sov. Phys. JETP} \textbf{\bibinfo{volume}{71}},
  \bibinfo{pages}{424} (\bibinfo{year}{1990}).

\bibitem[{\citenamefont{Afonin et~al.}(1988)}]{Afonin:1988gx}
\bibinfo{author}{\bibfnamefont{A.~I.} \bibnamefont{Afonin}}
  \bibnamefont{et~al.}, \bibinfo{journal}{Sov. Phys. JETP}
  \textbf{\bibinfo{volume}{67}}, \bibinfo{pages}{213} (\bibinfo{year}{1988}).

\bibitem[{\citenamefont{Greenwood et~al.}(1996)}]{Greenwood:1996pb}
\bibinfo{author}{\bibfnamefont{Z.~D.} \bibnamefont{Greenwood}}
  \bibnamefont{et~al.}, \bibinfo{journal}{Phys. Rev.}
  \textbf{\bibinfo{volume}{D53}}, \bibinfo{pages}{6054} (\bibinfo{year}{1996}).

\bibitem[{\citenamefont{Apollonio et~al.}(2003)}]{Apollonio:2002gd}
\bibinfo{author}{\bibfnamefont{M.}~\bibnamefont{Apollonio}}
  \bibnamefont{et~al.} (\bibinfo{collaboration}{CHOOZ}), \bibinfo{journal}{Eur.
  Phys. J.} \textbf{\bibinfo{volume}{C27}}, \bibinfo{pages}{331}
  (\bibinfo{year}{2003}), \eprint{hep-ex/0301017}.

\bibitem[{\citenamefont{Boehm et~al.}(2001)}]{Boehm:2001ik}
\bibinfo{author}{\bibfnamefont{F.}~\bibnamefont{Boehm}} \bibnamefont{et~al.}
  (\bibinfo{collaboration}{Palo Verde}), \bibinfo{journal}{Phys. Rev.}
  \textbf{\bibinfo{volume}{D64}}, \bibinfo{pages}{112001}
  (\bibinfo{year}{2001}), \eprint{hep-ex/0107009}.

\bibitem[{\citenamefont{Abe et~al.}(2014)}]{Abe:2014bwa}
\bibinfo{author}{\bibfnamefont{Y.}~\bibnamefont{Abe}} \bibnamefont{et~al.}
  (\bibinfo{collaboration}{Double Chooz}), \bibinfo{journal}{JHEP}
  \textbf{\bibinfo{volume}{1410}}, \bibinfo{pages}{86} (\bibinfo{year}{2014}),
  \eprint{arXiv:1406.7763}.

\bibitem[{\citenamefont{An et~al.}(2016)}]{An:2015nua}
\bibinfo{author}{\bibfnamefont{F.~P.} \bibnamefont{An}} \bibnamefont{et~al.}
  (\bibinfo{collaboration}{Daya Bay}), \bibinfo{journal}{Phys. Rev. Lett.}
  \textbf{\bibinfo{volume}{116}}, \bibinfo{pages}{061801}
  (\bibinfo{year}{2016}), \eprint{arXiv:1508.04233}.

\bibitem[{\citenamefont{Abazajian et~al.}(2012)}]{Abazajian:2012ys}
\bibinfo{author}{\bibfnamefont{K.~N.} \bibnamefont{Abazajian}}
  \bibnamefont{et~al.} (\bibinfo{year}{2012}), \eprint{arXiv:1204.5379}.

\bibitem[{\citenamefont{Frekers et~al.}(2011)\citenamefont{Frekers, Ejiri,
  Akimune, Adachi, Bilgier et~al.}}]{Frekers:2011zz}
\bibinfo{author}{\bibfnamefont{D.}~\bibnamefont{Frekers}},
  \bibinfo{author}{\bibfnamefont{H.}~\bibnamefont{Ejiri}},
  \bibinfo{author}{\bibfnamefont{H.}~\bibnamefont{Akimune}},
  \bibinfo{author}{\bibfnamefont{T.}~\bibnamefont{Adachi}},
  \bibinfo{author}{\bibfnamefont{B.}~\bibnamefont{Bilgier}},
  \bibnamefont{et~al.}, \bibinfo{journal}{Phys. Lett.}
  \textbf{\bibinfo{volume}{B706}}, \bibinfo{pages}{134} (\bibinfo{year}{2011}).

\bibitem[{\citenamefont{Giunti and Li}(2009)}]{Giunti:2009xz}
\bibinfo{author}{\bibfnamefont{C.}~\bibnamefont{Giunti}} \bibnamefont{and}
  \bibinfo{author}{\bibfnamefont{Y.}~\bibnamefont{Li}},
  \bibinfo{journal}{Phys.Rev.} \textbf{\bibinfo{volume}{D80}},
  \bibinfo{pages}{113007} (\bibinfo{year}{2009}), \eprint{arXiv:0910.5856}.

\bibitem[{\citenamefont{Palazzo}(2011)}]{Palazzo:2011rj}
\bibinfo{author}{\bibfnamefont{A.}~\bibnamefont{Palazzo}},
  \bibinfo{journal}{Phys. Rev.} \textbf{\bibinfo{volume}{D83}},
  \bibinfo{pages}{113013} (\bibinfo{year}{2011}), \eprint{arXiv:1105.1705}.

\bibitem[{\citenamefont{Palazzo}(2012)}]{Palazzo:2012yf}
\bibinfo{author}{\bibfnamefont{A.}~\bibnamefont{Palazzo}},
  \bibinfo{journal}{Phys. Rev.} \textbf{\bibinfo{volume}{D85}},
  \bibinfo{pages}{077301} (\bibinfo{year}{2012}), \eprint{arXiv:1201.4280}.

\bibitem[{\citenamefont{Palazzo}(2013{\natexlab{a}})}]{Palazzo:2013me}
\bibinfo{author}{\bibfnamefont{A.}~\bibnamefont{Palazzo}},
  \bibinfo{journal}{Mod.Phys.Lett.} \textbf{\bibinfo{volume}{A28}},
  \bibinfo{pages}{1330004} (\bibinfo{year}{2013}{\natexlab{a}}),
  \eprint{arXiv:1302.1102}.

\bibitem[{\citenamefont{Armbruster et~al.}(1998)}]{Armbruster:1998uk}
\bibinfo{author}{\bibfnamefont{B.}~\bibnamefont{Armbruster}}
  \bibnamefont{et~al.} (\bibinfo{collaboration}{KARMEN}),
  \bibinfo{journal}{Phys. Rev.} \textbf{\bibinfo{volume}{C57}},
  \bibinfo{pages}{3414} (\bibinfo{year}{1998}), \eprint{hep-ex/9801007}.

\bibitem[{\citenamefont{Auerbach et~al.}(2001)}]{Auerbach:2001hz}
\bibinfo{author}{\bibfnamefont{L.~B.} \bibnamefont{Auerbach}}
  \bibnamefont{et~al.} (\bibinfo{collaboration}{LSND}), \bibinfo{journal}{Phys.
  Rev.} \textbf{\bibinfo{volume}{C64}}, \bibinfo{pages}{065501}
  (\bibinfo{year}{2001}), \eprint{hep-ex/0105068}.

\bibitem[{\citenamefont{Conrad and Shaevitz}(2012)}]{Conrad:2011ce}
\bibinfo{author}{\bibfnamefont{J.}~\bibnamefont{Conrad}} \bibnamefont{and}
  \bibinfo{author}{\bibfnamefont{M.}~\bibnamefont{Shaevitz}},
  \bibinfo{journal}{Phys. Rev.} \textbf{\bibinfo{volume}{D85}},
  \bibinfo{pages}{013017} (\bibinfo{year}{2012}), \eprint{arXiv:1106.5552}.

\bibitem[{\citenamefont{Giunti and
  Laveder}(2011{\natexlab{b}})}]{Giunti:2011cp}
\bibinfo{author}{\bibfnamefont{C.}~\bibnamefont{Giunti}} \bibnamefont{and}
  \bibinfo{author}{\bibfnamefont{M.}~\bibnamefont{Laveder}},
  \bibinfo{journal}{Phys. Lett.} \textbf{\bibinfo{volume}{B706}},
  \bibinfo{pages}{200} (\bibinfo{year}{2011}{\natexlab{b}}),
  \eprint{arXiv:1111.1069}.

\bibitem[{\citenamefont{Dydak et~al.}(1984)}]{Dydak:1983zq}
\bibinfo{author}{\bibfnamefont{F.}~\bibnamefont{Dydak}} \bibnamefont{et~al.}
  (\bibinfo{collaboration}{CDHSW}), \bibinfo{journal}{Phys. Lett.}
  \textbf{\bibinfo{volume}{B134}}, \bibinfo{pages}{281} (\bibinfo{year}{1984}).

\bibitem[{\citenamefont{Maltoni and Schwetz}(2007)}]{Maltoni:2007zf}
\bibinfo{author}{\bibfnamefont{M.}~\bibnamefont{Maltoni}} \bibnamefont{and}
  \bibinfo{author}{\bibfnamefont{T.}~\bibnamefont{Schwetz}},
  \bibinfo{journal}{Phys. Rev.} \textbf{\bibinfo{volume}{D76}},
  \bibinfo{pages}{093005} (\bibinfo{year}{2007}), \eprint{arXiv:0705.0107}.

\bibitem[{\citenamefont{Hernandez and Smirnov}(2012)}]{Hernandez:2011rs}
\bibinfo{author}{\bibfnamefont{D.}~\bibnamefont{Hernandez}} \bibnamefont{and}
  \bibinfo{author}{\bibfnamefont{A.~Y.} \bibnamefont{Smirnov}},
  \bibinfo{journal}{Phys. Lett.} \textbf{\bibinfo{volume}{B706}},
  \bibinfo{pages}{360} (\bibinfo{year}{2012}), \eprint{arXiv:1105.5946}.

\bibitem[{\citenamefont{Giunti and
  Laveder}(2011{\natexlab{c}})}]{Giunti:2011hn}
\bibinfo{author}{\bibfnamefont{C.}~\bibnamefont{Giunti}} \bibnamefont{and}
  \bibinfo{author}{\bibfnamefont{M.}~\bibnamefont{Laveder}},
  \bibinfo{journal}{Phys.Rev.} \textbf{\bibinfo{volume}{D84}},
  \bibinfo{pages}{093006} (\bibinfo{year}{2011}{\natexlab{c}}),
  \eprint{arXiv:1109.4033}.

\bibitem[{\citenamefont{Adamson et~al.}(2011)}]{Adamson:2011ku}
\bibinfo{author}{\bibfnamefont{P.}~\bibnamefont{Adamson}} \bibnamefont{et~al.}
  (\bibinfo{collaboration}{MINOS}), \bibinfo{journal}{Phys. Rev. Lett.}
  \textbf{\bibinfo{volume}{107}}, \bibinfo{pages}{011802}
  (\bibinfo{year}{2011}), \eprint{arXiv:1104.3922}.

\bibitem[{\citenamefont{Mahn et~al.}(2012)}]{Mahn:2011ea}
\bibinfo{author}{\bibfnamefont{K.~B.~M.} \bibnamefont{Mahn}}
  \bibnamefont{et~al.} (\bibinfo{collaboration}{SciBooNE-MiniBooNE}),
  \bibinfo{journal}{Phys. Rev.} \textbf{\bibinfo{volume}{D85}},
  \bibinfo{pages}{032007} (\bibinfo{year}{2012}), \eprint{arXiv:1106.5685}.

\bibitem[{\citenamefont{Cheng et~al.}(2012)}]{Cheng:2012yy}
\bibinfo{author}{\bibfnamefont{G.}~\bibnamefont{Cheng}} \bibnamefont{et~al.}
  (\bibinfo{collaboration}{SciBooNE-MiniBooNE}), \bibinfo{journal}{Phys. Rev.}
  \textbf{\bibinfo{volume}{D86}}, \bibinfo{pages}{052009}
  (\bibinfo{year}{2012}), \eprint{arXiv:1208.0322}.

\bibitem[{\citenamefont{Aartsen et~al.}(2016)}]{TheIceCube:2016oqi}
\bibinfo{author}{\bibfnamefont{M.~G.} \bibnamefont{Aartsen}}
  \bibnamefont{et~al.} (\bibinfo{collaboration}{IceCube}),
  \bibinfo{journal}{Phys. Rev. Lett.} \textbf{\bibinfo{volume}{117}},
  \bibinfo{pages}{071801} (\bibinfo{year}{2016}), \eprint{1605.01990}.

\bibitem[{\citenamefont{Huber et~al.}(2005)\citenamefont{Huber, Lindner, and
  Winter}}]{globes_1}
\bibinfo{author}{\bibfnamefont{P.}~\bibnamefont{Huber}},
  \bibinfo{author}{\bibfnamefont{M.}~\bibnamefont{Lindner}}, \bibnamefont{and}
  \bibinfo{author}{\bibfnamefont{W.}~\bibnamefont{Winter}},
  \bibinfo{journal}{Comput. Phys. Commun.} \textbf{\bibinfo{volume}{167}},
  \bibinfo{pages}{195} (\bibinfo{year}{2005}), \eprint{hep-ph/0407333}.

\bibitem[{\citenamefont{Huber et~al.}(2007)\citenamefont{Huber, Kopp, Lindner,
  Rolinec, and Winter}}]{globes_2}
\bibinfo{author}{\bibfnamefont{P.}~\bibnamefont{Huber}},
  \bibinfo{author}{\bibfnamefont{J.}~\bibnamefont{Kopp}},
  \bibinfo{author}{\bibfnamefont{M.}~\bibnamefont{Lindner}},
  \bibinfo{author}{\bibfnamefont{M.}~\bibnamefont{Rolinec}}, \bibnamefont{and}
  \bibinfo{author}{\bibfnamefont{W.}~\bibnamefont{Winter}},
  \bibinfo{journal}{Comput. Phys. Commun.} \textbf{\bibinfo{volume}{177}},
  \bibinfo{pages}{432} (\bibinfo{year}{2007}), \eprint{hep-ph/0701187}.

\bibitem[{\citenamefont{Abe et~al.}(2013)}]{t2k_flux}
\bibinfo{author}{\bibfnamefont{K.}~\bibnamefont{Abe}} \bibnamefont{et~al.}
  (\bibinfo{collaboration}{T2K}), \bibinfo{journal}{Phys. Rev.}
  \textbf{\bibinfo{volume}{D87}}, \bibinfo{pages}{012001}
  (\bibinfo{year}{2013}), \bibinfo{note}{[Addendum: Phys.
  Rev.D87,no.1,019902(2013)]}, \eprint{1211.0469}.

\bibitem[{\citenamefont{Ravonel~Salzgeber}(2015)}]{t2k_antinu_flux}
\bibinfo{author}{\bibfnamefont{M.}~\bibnamefont{Ravonel~Salzgeber}}
  (\bibinfo{collaboration}{T2K}) (\bibinfo{year}{2015}), \eprint{1508.06153}.

\bibitem[{\citenamefont{Rocco}(2016)}]{nova_thesis}
\bibinfo{author}{\bibfnamefont{D.~R.} \bibnamefont{Rocco}}, Ph.D. thesis,
  \bibinfo{school}{Minnesota U.} (\bibinfo{year}{2016}),
  \urlprefix\url{http://lss.fnal.gov/archive/thesis/2000/fermilab-thesis-2016-15.pdf}.

\bibitem[{\citenamefont{Palazzo}(2013{\natexlab{b}})}]{palazzo_reactors}
\bibinfo{author}{\bibfnamefont{A.}~\bibnamefont{Palazzo}},
  \bibinfo{journal}{JHEP} \textbf{\bibinfo{volume}{10}}, \bibinfo{pages}{172}
  (\bibinfo{year}{2013}{\natexlab{b}}), \eprint{1308.5880}.

\bibitem[{\citenamefont{Yu}(2016)}]{daya_bay_neutrino_2016}
\bibinfo{author}{\bibfnamefont{Z.}~\bibnamefont{Yu}} (\bibinfo{year}{2016}),
  \bibinfo{note}{presented at the 27th International Conference on Neutrino
  Physics and Astrophysics in July 2016 in London,}.

\bibitem[{\citenamefont{Joo}(2016)}]{reno_neutrino_2016}
\bibinfo{author}{\bibfnamefont{K.~K.} \bibnamefont{Joo}}
  (\bibinfo{year}{2016}), \bibinfo{note}{presented at the 27th International
  Conference on Neutrino Physics and Astrophysics in July 2016 in London,}.

\bibitem[{\citenamefont{Maltoni and Schwetz}(2003)}]{Maltoni:2003cu}
\bibinfo{author}{\bibfnamefont{M.}~\bibnamefont{Maltoni}} \bibnamefont{and}
  \bibinfo{author}{\bibfnamefont{T.}~\bibnamefont{Schwetz}},
  \bibinfo{journal}{Phys. Rev.} \textbf{\bibinfo{volume}{D68}},
  \bibinfo{pages}{033020} (\bibinfo{year}{2003}), \eprint{hep-ph/0304176}.

\bibitem[{\citenamefont{Hannestad et~al.}(2012)\citenamefont{Hannestad,
  Tamborra, and Tram}}]{Hannestad:2012ky}
\bibinfo{author}{\bibfnamefont{S.}~\bibnamefont{Hannestad}},
  \bibinfo{author}{\bibfnamefont{I.}~\bibnamefont{Tamborra}}, \bibnamefont{and}
  \bibinfo{author}{\bibfnamefont{T.}~\bibnamefont{Tram}},
  \bibinfo{journal}{JCAP} \textbf{\bibinfo{volume}{1207}}, \bibinfo{pages}{025}
  (\bibinfo{year}{2012}), \eprint{arXiv:1204.5861}.

\bibitem[{\citenamefont{Dasgupta and Kopp}(2014)}]{Dasgupta:2013zpn}
\bibinfo{author}{\bibfnamefont{B.}~\bibnamefont{Dasgupta}} \bibnamefont{and}
  \bibinfo{author}{\bibfnamefont{J.}~\bibnamefont{Kopp}},
  \bibinfo{journal}{Phys.Rev.Lett.} \textbf{\bibinfo{volume}{112}},
  \bibinfo{pages}{031803} (\bibinfo{year}{2014}), \eprint{arXiv:1310.6337}.

\bibitem[{\citenamefont{Hannestad et~al.}(2014)\citenamefont{Hannestad, Hansen,
  and Tram}}]{Hannestad:2013ana}
\bibinfo{author}{\bibfnamefont{S.}~\bibnamefont{Hannestad}},
  \bibinfo{author}{\bibfnamefont{R.~S.} \bibnamefont{Hansen}},
  \bibnamefont{and} \bibinfo{author}{\bibfnamefont{T.}~\bibnamefont{Tram}},
  \bibinfo{journal}{Phys.Rev.Lett.} \textbf{\bibinfo{volume}{112}},
  \bibinfo{pages}{031802} (\bibinfo{year}{2014}), \eprint{arXiv:1310.5926}.

\bibitem[{\citenamefont{Saviano et~al.}(2013)}]{Saviano:2013ktj}
\bibinfo{author}{\bibfnamefont{N.}~\bibnamefont{Saviano}} \bibnamefont{et~al.},
  \bibinfo{journal}{Phys.Rev.} \textbf{\bibinfo{volume}{D87}},
  \bibinfo{pages}{073006} (\bibinfo{year}{2013}), \eprint{arXiv:1302.1200}.

\bibitem[{\citenamefont{Archidiacono et~al.}(2016)\citenamefont{Archidiacono,
  Gariazzo, Giunti, Hannestad, Hansen, Laveder, and
  Tram}}]{Archidiacono:2016kkh}
\bibinfo{author}{\bibfnamefont{M.}~\bibnamefont{Archidiacono}},
  \bibinfo{author}{\bibfnamefont{S.}~\bibnamefont{Gariazzo}},
  \bibinfo{author}{\bibfnamefont{C.}~\bibnamefont{Giunti}},
  \bibinfo{author}{\bibfnamefont{S.}~\bibnamefont{Hannestad}},
  \bibinfo{author}{\bibfnamefont{R.}~\bibnamefont{Hansen}},
  \bibinfo{author}{\bibfnamefont{M.}~\bibnamefont{Laveder}}, \bibnamefont{and}
  \bibinfo{author}{\bibfnamefont{T.}~\bibnamefont{Tram}},
  \bibinfo{journal}{JCAP} \textbf{\bibinfo{volume}{1608}}, \bibinfo{pages}{067}
  (\bibinfo{year}{2016}), \eprint{1606.07673}.

\bibitem[{\citenamefont{Agarwalla et~al.}(2017)\citenamefont{Agarwalla,
  Chatterjee, and Palazzo}}]{Agarwalla:2016xlg}
\bibinfo{author}{\bibfnamefont{S.~K.} \bibnamefont{Agarwalla}},
  \bibinfo{author}{\bibfnamefont{S.~S.} \bibnamefont{Chatterjee}},
  \bibnamefont{and} \bibinfo{author}{\bibfnamefont{A.}~\bibnamefont{Palazzo}},
  \bibinfo{journal}{Phys. Rev. Lett.} \textbf{\bibinfo{volume}{118}},
  \bibinfo{pages}{031804} (\bibinfo{year}{2017}), \eprint{1605.04299}.

\end{thebibliography}

\end{document}